\documentclass[final,3p,times]{elsarticle}
\usepackage{graphicx} 

\usepackage{amssymb}
\usepackage{amsmath}
\usepackage[table,xcdraw]{xcolor}
\usepackage[ruled,vlined]{algorithm2e}

\usepackage{bm}

\usepackage{url}
\usepackage{hyperref}

\usepackage[english]{babel}
\usepackage{multirow}
\usepackage{booktabs}
\hypersetup{
    citecolor=red,
    colorlinks=true,
    linkcolor=blue,
    filecolor=magenta,      
    urlcolor=cyan,
    }

\usepackage{float}

\newcommand{\beginsupplement}{%

        \setcounter{table}{0}
        \renewcommand{\thetable}{S\arabic{table}}%
        \setcounter{figure}{0}
        \renewcommand{\thefigure}{S\arabic{figure}}
        \setcounter{equation}{0}
        \renewcommand{\theequation}{S\arabic{equation}}%
        \setcounter{page}{1}
        \renewcommand{\thepage}{S\arabic{page}} 
  
        \setlength{\parskip}{5pt}
        \everypar{\setlength{\parindent}{24pt}}

        \setcounter{section}{0}%
        \renewcommand{\thesection}{S\arabic{section}}%

      \renewcommand{\thesubsection}{\thesection.\arabic{subsection}}%

       \hypersetup{
            colorlinks = true,
            linkbordercolor = {blue},
            }
}

\journal{Computer Physics Communications}

\title{Active learning of collinear magnetic Moment Tensor Potentials using the spin-MLIP package from soft-constrained spin-polarized DFT calculations: a case study of Fe--Pd}

\author[SKOLT]{Arseniy Burov}
 \author[SKOLT,HSE]{Alexey S. Kotykhov}
    \author[SKOLT]{Dmitry A. Aksyonov}
    \author[HSE]{Ivan S. Novikov\corref{cor}}
    \author[CALTECH]{Vladimir V. Ladygin}
    
    \cortext[cor]{isnovikov@hse.ru}
    
    \affiliation[SKOLT]{organization={Skolkovo Institute of Science and Technology},
        addressline={Bolshoy boulevard 30}, 
        city={Moscow},
        postcode={121205}, 
        country={Russian Federation}}

    \affiliation[HSE]{organization={HSE University, Faculty of Computer Science},
        addressline={Pokrovsky boulevard 11},
        city={Moscow},
        postcode={109028}, 
        country={Russian Federation}}

    \affiliation[CALTECH]{organization={California Institute of Technology},
    addressline={1200 E. California boulevard}, 
    city={Pasadena},
    state={California},
    postcode={91125},
    country={United States of America}}

\begin{document}

\begin{frontmatter}

    \begin{abstract}
    Explicit incorporation of magnetic degrees of freedom in machine-learning interatomic potentials (magnetic MLIPs) plays a crucial role in the correct description of magnetic materials and their properties. An important ingredient for fitting of magnetic MLIPs is spin-polarized density functional theory (DFT) calculations with non-equilibrium magnetic moments, i.e. DFT calculations with constraints on magnetic moments. In this study, we present a workflow for active learning of magnetic Moment Tensor Potential (mMTP) during molecular dynamics (MD) simulations. Magnetic MTP and its active learning algorithm were implemented in the open-source spin-MLIP code, DFT soft-constrained spin-polarized calculations were performed with the VASP code, and MD simulations were conducted in the open-source LAMMPS code. We test our workflow on the Fe--Pd crystal. The dependencies of magnetization and phonon density of states (DOSs) on the volume of a supercell (or, pressure) are in good agreement with those calculated with DFT. Furthermore, the calculated DOSs correspond to the experimental ones.
    \end{abstract}

\end{frontmatter}

\section{Introduction}

An accurate description of magnetism plays an important role in the theoretical study of magnetic materials and their properties. Nowadays, machine-learning interatomic potentials (MLIPs) are a popular and widely-used tool for the theoretical investigation of materials. During the last five years, many MLIPs explicitly incorporating magnetic degrees of freedom (magnetic MLIPs) have been developed~\cite{eckhoff2021_mhdnnp, chapman2022_nn_defect_Fe, nikolov2021_snap_heisenberg, domina2022_spinSNAP, yu2022_spinGNN, novikov2022magnetic, yu2024_spinGNN++, yuan2024_magENN, yang2024deep, fan2024coarsening, rinaldi2024non, gao2024machine}. Magnetic MLIPs are typically trained on energies, forces, and stresses calculated with density functional theory, yield the accuracy of DFT calculations, and enable predicting magnetic properties of materials such as Curie temperature \cite{chapman2022_nn_defect_Fe, nikolov2021_snap_heisenberg, yu2024_spinGNN++, rinaldi2024non}, N\'eel temperature \cite{eckhoff2021_mhdnnp, yu2024_spinGNN++}, and magnetization \cite{chapman2022_nn_defect_Fe, nikolov2021_snap_heisenberg, rinaldi2024non}, as well as vibrational and mechanical properties like phonon spectra \cite{rinaldi2024non, yang2024deep}, bulk modulus \cite{nikolov2021_snap_heisenberg}, and elastic constants \cite{nikolov2021_snap_heisenberg, rinaldi2024non} for accurate prediction of which it is necessary to take magnetism into account in magnetic materials \cite{yildirim2009frustrated,kormann2012atomic}.

The main difficulty in the construction of magnetic MLIPs is the creation of a training set for their fitting. The first challenge here is the automated selection of the configurations to be added to a training set. As opposed to the non-magnetic case, here we also have magnetic moments as an additional degree of freedom and, therefore, it becomes even more difficult to create a training set manually by following only physical intuition. Furthermore, an algorithm for automated selection of configurations enables reducing the number of configurations in a training set, which is critical, as spin-polarized DFT calculations typically used to create a training set for magnetic MLIPs are even more time consuming than conventional (non-spin-polarized) DFT calculations. The second challenge is the need to have configurations with non-equilibrium (excited) magnetic moments in a training set, as they are explicitly included in a functional form of magnetic MLIPs along with atomic positions, types, and lattice vectors. Just as we require non-equilibrium atomic positions to properly fit non-magnetic MLIPs, we also need non-equilibrium magnetic moments to fit magnetic MLIPs. Due to the above reasons, it can be rather difficult for a user to implement a workflow that enables automated training of magnetic MLIPs. Furthermore, to our knowledge, only the codes DeePSPIN~\cite{DeepSPIN} and mag-ace~\cite{magACE} include magnetic MLIPs and are open-source.

In this study, we describe a workflow for the automated construction of training sets to fit the magnetic MLIP during atomistic simulations --- specifically, magnetic Moment Tensor Potential (magnetic MTP, or mMTP) proposed in~\cite{novikov2022magnetic} for single-component magnetic materials and in~\cite{Kotykhov2023-cDFT-mMTP} for multi-component magnetic materials and provide a workflow for its automated fitting. As an example of atomistic simulations, we consider molecular dynamics simulations. To select the configurations to be added to a training set, we use an active learning algorithm formulated in~\cite{kotykhov2025_CrN} and based on the maxvol algorithm~\cite{goreinov2010_maxvol}. To compute DFT energies, forces, and stresses for selected configurations, we utilize soft-constrained spin-polarized DFT calculations that enable approximately preserving non-equilibrium magnetic moments as a result of calculations. In addition to the algorithm for automated fitting of mMTP, we introduce an open-source code called spin-MLIP in which mMTP and the algorithm for its active learning were implemented~\cite{spin-MLIP}, its interface~\cite{spin-MLIP-LAMMPS} with the LAMMPS code~\cite{plimpton1995fast,thompson2022lammps}, and the workflow~\cite{pipeline} including scripts for molecular dynamics simulations with LAMMPS, for selection of configurations with spin-MLIP, for soft-constrained DFT calculations with the VASP code~\cite{Kresse:1996,Kresse:1999}, and for fitting of mMTP also using spin-MLIP. As an example of an atomic system, we take Fe--Pd and calculate the dependencies of its magnetization and density of states (DOSs) on the volume of a supercell (or, pressure) with the fitted mMTP. We chose Fe--Pd as the test system because it is a prototypical case of strong magnetovolume coupling at finite temperature. Fe-rich face-centered cubic (fcc) Fe--Pd alloys belong to the family of Fe-based Invar alloys, and the magnetization of $\text{Fe}_{0.68}\text{Pd}_{0.32}$ decreases markedly under pressure \cite{matsushita2002anomalous, priesen2025thermodynamic}. Fe--Pd is also a stringent test of the soft-constrained DFT scheme, because the constraints must be satisfied simultaneously for large Fe moments and for induced Pd moments an order of magnitude smaller. Finally, Ref.~\cite{priesen2025thermodynamic} reports, for the same composition, the pressure-volume relation, the Fe magnetization, and the room-temperature Fe-partial phonon DOS at several pressures, enabling a direct validation of mMTP against both DFT and experiment. The corresponding dependencies calculated with mMTP are close to the DFT ones, while the phonon DOSs are also in good agreement with the experimental data.

Without loss of generality, the presented workflow can be applied to other atomic systems. For example, the idea of training on configurations with non-equilibrium magnetic moments has previously been demonstrated for the Fe--Al system \cite{Kotykhov2023-cDFT-mMTP}, while the active learning methodology for mMTPs was also applied to CrN \cite{kotykhov2025_CrN}. However, previous studies employed hard-constrained spin-polarized DFT calculations, which enable accurate preservation of non-equilibrium magnetic moments after computations, whereas the present work uses soft-constrained DFT calculations. As demonstrated in this study, soft-constrained DFT calculations are sufficient to preserve non-equilibrium magnetic moments of individual atoms with the accuracy required for reliable mMTP training.

\section{Methodology}

\subsection{Magnetic Moment Tensor Potential}

We used magnetic Moment Tensor Potential (magnetic MTP, mMTP) as a machine-learning interatomic potential for magnetic materials. The mMTP model was first developed for single-component materials in~\cite{novikov2022magnetic} and is based on a non-magnetic version of MTP~\cite{shapeev2016moment}. The functional form of the chosen potential explicitly depends on the magnitude of the scalar magnetic moments of the atoms, thus enabling the potential to describe collinear magnetism not only qualitatively but also quantitatively. 

The mMTP model is local, i.e. the potential energy of a configuration is decomposed into the sum of the energies of each atomic environment $\mathcal{N}_{i}$:

\begin{equation}
\label{eq:total_energy}
E^{\rm mMTP} = \sum_{i=1}^{N_{\text{atoms}}} V(\mathcal{N}_{i}),
\end{equation}
where $N_{\text{atoms}}$ is the number of atoms in a supercell and an atomic neighborhood $\mathcal{N}_{i}$ is a set of atoms within a sphere of cutoff radius $R_{\text{cut}}$ around the atom $i$. Each atomic neighborhood contains information about the distances $|\textbf{r}_{ij}|$ between the central atom and its neighbors, as well as the atomic types $z_i$ and $z_j$, and the magnetic moments $m_i$ and $m_j$ of central and neighboring atoms:

\begin{equation}
\label{eq:nbh}
\mathcal{N}_{i} = \{ (\textbf{r}_{ij}, z_{i}, z_{j}, {\mathrm m}_{i}, {\mathrm m}_{j}), j = 1, \dots, N^{i}_{\text{nbh}}: |\textbf{r}_{ij}| < R_{\text{cut}}\}.
\end{equation}

The energy for each atomic environment is represented as a linear combination of basis functions $B_{\alpha}$:

\begin{equation}
\label{eq:nbh_energy}
V(\mathcal{N}_{i}) = \sum_{\alpha=1}^{N_{\rm lin}} \xi_{\alpha} B_{\alpha}.
\end{equation}

The coefficients $\xi_{\alpha}$ are linear trainable parameters, $N_{\rm lin}$ is the number of these coefficients. The basis functions are obtained as contractions of moment tensor descriptors $M_{\mu, \nu}$ yielding a scalar:

\begin{equation}
\label{eq:moment_descriptor}
M_{\mu,\nu}(\mathcal{N}_i)=\sum_{j} f_{\mu}(|{\textbf{r}}_{ij}|,z_i,z_j,{\mathrm m}_{i}, {\mathrm m}_{j}) {\textbf{r}}_{ij}^{\otimes \nu}.
\end{equation}

The function $f_{\mu}$ depends on the interatomic distances and magnetic moments of the particles for each pair of atomic types and forms the radial part. The outer product $\otimes$, taken $\nu$ times, represents the angular part of the moment tensor descriptor. Thus, when $\nu$ is equal to zero, the angular part will be a scalar. When $\nu$ is equal to one, the angular part represents the initial vector of the interatomic distance, and when $\nu$ is equal to two, it will be a matrix.

The radial part of the moment tensor descriptor consists of various polynomials:

\begin{equation}
\label{eq:radial_part}
f_{\mu}(| {\textbf{r}}_{ij}|,z_i,z_j,{\mathrm m}_i,{\mathrm m}_j) = s \sum_{\zeta=1}^{N_{\phi}} \sum_{\beta=1}^{N_{\psi}}\sum_{\gamma=1}^{N_{\psi}} c_{\mu,z_i,z_j}^{\zeta,\beta,\gamma} \phi_{\zeta}(|\textbf{r}_{ij}|) \psi_{\beta}({\mathrm m}_i)\psi_{\gamma}({\mathrm m}_j) (R_{\rm cut} - |\textbf{r}_{ij}|)^2.
\end{equation}

The functions $\phi_{\zeta}$, $\psi_{\beta}$, and $\psi_{\gamma}$ are Chebyshev polynomials of the first kind corresponding to the orders indicated by the subscript. Symbols $N_{\phi}$ and $N_{\psi}$ denote the sizes of the radial and magnetic bases, respectively. Symbol $s$ denotes the scaling parameter (a positive number). The details of scaling are given in Supplementary Information. The interatomic distance is defined on the interval $(R_{\rm min},R_{\rm cut})$ and the magnetic moments from the interval $(-M_{\rm max}^{z_i},M_{\rm max}^{z_i})$ are mapped onto the interval $(-1, 1)$. Each type of atom $z_i$ has its own maximum magnetic moment $M_{\rm max}^{z_i}$ in the training set. The parameters $c_{\mu,z_i,z_j}^{\zeta,\beta,\gamma}$ are trainable. The inclusion of the functions $\psi_{\beta}$ and $\psi_{\gamma}$ is the main difference between mMTP and the original non-magnetic MTP. Because of this, the radial part of the mMTP contains $N_{\psi}^{2}$ times more parameters than the original MTP. Since the functions $\psi_{\beta}$ and $\psi_{\gamma}$ depend only on the magnitude of magnetic moments and the angular part does not include magnetic moments at all, mMTP can only describe collinear magnetism.

As an infinite number of contractions of moment tensor descriptors yielding a scalar can be constructed for different values of $\mu$ and $\nu$, it is necessary to limit a number of contractions. Therefore, the concept of moment tensor descriptor level is introduced:

\begin{equation}
\label{eq:level}
\text{lev} M_{\mu, \nu} = 2 + 4 \mu + \nu.
\end{equation}

The coefficients in Eq.~\eqref{eq:level} were empirically selected in~\cite{gubaev2019accelerating}. The level of contraction of moments is defined as the sum of the levels of moments. For example, $\text{lev} (M_{0, 0}^4)$ = 8 and  $\text{lev} (M_{0, 1} \cdot M_{2, 1})$ = 14. The concept of a potential level $\text{lev}_{\text{mMTP}}$ is also introduced, representing the maximum permissible level of moment contractions. Thus, basis functions consist of all permitted moment contractions, i.e., whose level is not greater than the potential level ($\text{lev} B_{\alpha} \leq \text{lev}_{\text{mMTP}}$). We also note that $\text{lev}_{\text{mMTP}}$ determines the number of linear parameters $N_{\rm lin}$ which corresponds to the number of the basis functions $B_{\alpha}$ included in concrete MTP functional form.

Denote $\bm \theta = \{ \xi_{\alpha}, c_{\mu,z_i,z_j}^{\zeta,\beta,\gamma} \}$ by the set of all trainable parameters from Eqs.~\eqref{eq:nbh_energy} and~\eqref{eq:radial_part}. The symbols $\mathbf{R}$, $Z$, $M$, and $L$ denote the coordinates, types, and magnetic moments of atoms in a system, and lattice vectors, respectively. For physical reasons, the energy of the system should not change when the magnetic moments of all atoms are flipped. To satisfy this condition, the functional form of the potential is symmetrized:

\begin{equation}
\label{eq:symmetrization}
E^{\rm mMTP}(\bm \theta) = \dfrac{E(\bm \theta, \mathbf{R}, Z, L, M)+E(\bm \theta, \mathbf{R}, Z, L, -M)}{2}.
\end{equation}

The optimal parameters of the potential are determined by minimizing the loss function:

\begin{equation}
\label{eq:loss}
\text{Loss}(\bm \theta) =  \sum_{k=1}^{K} \Biggl[ w_e \Bigl(E^{\rm mMTP}_k(\bm \theta) - E_k^{\rm DFT} \Bigr)^2 + w_f \sum_{i=1}^{N_{\text{atoms}}} \Bigl(\textbf{f}_{i,k}^{\rm mMTP}( \bm \theta) - \textbf{f}_{i,k}^{\rm DFT} \Bigr)^2  + w_s \sum_{a, b = 1}^{3} \Bigl(\sigma_{ab,k}^{\rm mMTP}( \bm \theta) - \sigma_{ab,k}^{\rm DFT} \Bigr)^2  \Biggr].
\end{equation}

Thus, the training set for the potential consists of configurations with coordinates, types and magnetic moments of atoms together with the corresponding energies $E_k^{\rm DFT}$, forces $\textbf{f}_{i,k}^{\rm DFT}$ and stresses $\sigma_{ab,k}^{\rm DFT}$ calculated with DFT for each configuration $k$. The weights $w_{e}$, $w_{f}$, and $w_{s}$ in the loss function are non-negative numbers that control the relative importance of training on energies, forces, and stresses, respectively. Further details on the determination of the relative loss function weights are provided in \cite{novoselov2019moment}. To minimize the loss function, the Broyden-Fletcher-Goldfarb-Shanno (BFGS) algorithm~\cite{broyden1970convergence, fletcher1970new, goldfarb1970family, shanno1970conditioning} with a random initial guess was used. To speed up convergence of the BFGS algorithm, scaling of mMTP was also utilized (see Supplementary Information for details).

During simulations using mMTP, prior to each molecular dynamics or geometry optimization step, equilibrium magnetic moments are obtained by minimizing the energy as a function of the atomic magnetic moments. Consequently, for robust simulations, mMTP should accurately describe the potential energy surface for non-equilibrium magnetic moments in the vicinity of the equilibrium ones. Therefore, it is particularly important to include configurations with non-equilibrium magnetic moments in the training set and, consequently, to have the capability to perform constrained spin-polarized DFT calculations.

\subsection{Active learning algorithm}

To automate the process of constructing a training set, one can use an active learning algorithm, which enables selecting the most relevant configurations during atomic simulations for subsequent recalculation using DFT and adding the selected configurations to the training set. For non-magnetic MTP the algorithm was formulated in~\cite{podryabinkin2017active, gubaev2018machine}, and for mMTP in~\cite{kotykhov2025_CrN}.

Assuming that we have a vector of optimal trainable parameters $\bar{\bf \theta}$ after training only on energies, we can linearize each term in the loss function:

\begin{equation}
\label{eq:linearization}
E_{k}^{\text{DFT}} - E_{k}^{\text{mMTP}}(\bm \theta) \approx E_{k}^{\text{DFT}} - \sum_{p=1}^{P} \frac{\partial E_{k}^{\text{mMTP}}(\bar{\bm \theta})}{\partial \theta_p} ( \theta_p - \bar{\theta}_p).
\end{equation}

After this, one can think of learning the potential as solving a system of equations for $\bm \theta$:

\begin{equation}
\label{eq:system}
\sum_{p=1}^{P} \theta_{p} \frac{\partial E_{k}^{\text{mMTP}}(\bar{\bm \theta})}{\partial \theta_{p}} = E_{k}^{\text{DFT}} + \sum_{p=1}^{P} \bar{\theta}_{p} \frac{\partial E_{k}^{\text{mMTP}}(\bar{\bm \theta})}{\partial \theta_{p}}.
\end{equation}

Thus, the matrix of the system of equations has the shape $K \times P$, where $K$ denotes the number of configurations ${\text{cfg} }^{(1)}, \ldots, {\text{cfg} }^{(K)}$ in the training set and $P$ is the number of trainable parameters:

\begin{equation}
\label{eq:active_set}
B = \begin{pmatrix}
\frac{\partial E^{\rm mMTP}}{\partial \theta_1} ( {\bar{\bm \theta}}, { \text{cfg} }^{(1)}) & \dots & \frac{\partial E^{\rm mMTP}}{\partial \theta_P} ( {\bar{\bm \theta}}, { \text{cfg} }^{(1)}) \\
\vdots & \ddots & \vdots \\
\frac{\partial E^{\rm mMTP}}{\partial \theta_1} ( {\bar{\bm \theta}}, { \text{cfg} }^{(K)}) & \dots & \frac{\partial E^{\rm mMTP}}{\partial \theta_P} ( {\bar{\bm \theta}}, { \text{cfg} }^{(K)})
\end{pmatrix}.
\end{equation}

After this, it is possible to select a submatrix $A$ of size $P \times P$ with the maximum absolute value of determinant from the matrix $B$ using the maxvol algorithm~\cite{goreinov2010_maxvol}. This can be interpreted as searching for the most diverse configurations from the training set.

After constructing the matrix A, an extrapolation grade $\gamma(\text{cfg}^{*})$ of each encountered configuration $\text{cfg}^{*}$ can be assessed during atomic simulations:

\begin{equation}
\label{eq:extrapolation_grade}
\gamma(\text{cfg}^{*}) = \max\limits_{1 \le j \le P} |c_j|.
\end{equation}

The vector $\textbf{c} = (c_1, \ldots, c_{P})$ is defined as:

\begin{equation}
\label{eq:c}
\textbf{c} = \Bigl(\frac{\partial E^{\rm mMTP}}{\partial \theta_1} ( {\bar{\bm \theta}}, { \text{cfg} }^*), \dots, \frac{\partial E^{\rm mMTP}}{\partial \theta_P} ( {\bar{\bm \theta}}, { \text{cfg} }^*) \Bigr) \  A^{-1}.
\end{equation}

The extrapolation grade for configuration $\text{cfg}^{*}$ can be interpreted as an indicator of how much the absolute value of the determinant of matrix $A$ grows when configuration $\text{cfg}^{*}$ is added to the training set, i.e. how much the diversity of the training set increases. When $\gamma(\text{cfg}^{*}) < 1$, it is considered that the potential interpolates the energy value of the configuration $\text{cfg}^{*}$, and when $\gamma(\text{cfg}^{*}) > 1$, on the contrary, it extrapolates.

To avoid selecting configurations with an extrapolation grade that is too high or too low, a two-threshold scheme is used, in which configurations with $\gamma(\text{cfg}^{*}) > \gamma_{\text{low}}$ are preselected and the preselection (or, extrapolation control) stops completely if a configuration with $\gamma(\text{cfg}^{*}) > \gamma_{\text{up}}$ is encountered. After the simulation is terminated, the matrix $A$ is updated using the maxvol algorithm with configurations from the set of selected configurations. For new configurations included in the matrix $A$, the energies, forces, and stresses are calculated using DFT, after which these configurations are added to the training set. 

For the active learning workflow to work correctly, the configurations calculated using DFT should correspond as closely as possible to the selected ones. Therefore, constraints on the magnetic moments should be satisfied with the highest possible accuracy during DFT calculations. To that end, we use soft-constrained DFT calculations (see subsection~\ref{SoftConstrainedDFT_main}) and add the calculated configurations to the training set. The potential is further trained on the updated training set, after which a new iteration of active learning begins. The active learning algorithm finishes when the simulation does not encounter a single configuration with an extrapolation grade higher than $\gamma_{\text{low}}$.

\subsection{Soft-constrained spin-polarized density functional theory calculations} \label{SoftConstrainedDFT_main}

In conventional (ground-state) spin-polarized Kohn--Sham DFT, we minimize the Hohenberg--Kohn total energy functional $E[\rho; \mathbf{R}, Z, L]$ with respect to the electron density $\rho = \rho(\mathbf{r})$, which is constructed from the Kohn--Sham orbitals $\{\psi_i^{\sigma}\}$ and their occupation numbers $f_{i\sigma}$~\cite{martin2020electronic}:

\begin{equation}
\label{eq:el_density}
\rho(\mathbf{r}) = \sum_{i} f_{i\uparrow} |\psi_i^{\uparrow}(\mathbf{r})|^2
+ \sum_{i} f_{i\downarrow} |\psi_i^{\downarrow}(\mathbf{r})|^2
= \rho^{\uparrow}(\mathbf{r}) + \rho^{\downarrow}(\mathbf{r}),
\end{equation}
where $\psi_i^{\uparrow}(\mathbf{r})$ and $\psi_i^{\downarrow}(\mathbf{r})$ are the Kohn--Sham orbitals corresponding to spin-up and spin-down electrons, respectively, and $f_{i\sigma} \in [0, 1]$ are their occupation numbers ($\sigma = \uparrow, \downarrow$). By solving the following minimization problem:

\begin{equation} \label{eq:dft_gs_en}
E_{\text{opt}}^{\rm DFT} \equiv E_{\text{opt}}^{\rm DFT}(\mathbf{R}, Z, L) = \min_{\rho} E[\rho; \mathbf{R}, Z, L],
\end{equation}
we obtain the optimal electron density $\rho_{\rm opt} = \rho^{\uparrow}_{\rm opt} + \rho^{\downarrow}_{\rm opt}$ and the optimal energy $E_{\text{opt}}^{\rm DFT}$. We note that positions of nuclei $\mathbf{R}$, lattice vectors $L$, and atomic types $Z$ are fixed when solving Eq.~\eqref{eq:dft_gs_en}. Using the optimal electron densities $\rho^{\uparrow}_{\mathrm{opt}}$ and $\rho^{\downarrow}_{\mathrm{opt}}$, the magnetic moment of the $i$-th atom can be calculated by integrating over a specific sphere $\Omega_i$ around the $i$-th atom:

\begin{equation} \label{eq:opt_magmom}
\mathrm{m}_i^{\rm opt} = \int_{\Omega_i} {\rho}^{\mathrm{mag}}_{\rm opt}(\mathbf{r}) d\mathbf{r} \equiv \int_{\Omega_i} \left(\rho^{\uparrow}_{\rm opt}(\mathbf{r}) - \rho^{\downarrow}_{\rm opt}(\mathbf{r})\right) d\mathbf{r}.
\end{equation}

In the case of conventional spin-polarized DFT calculations, the resulting magnetic moments $\mathrm{m}_i^{\rm opt}$, $i = 1, \ldots, N_{\rm atoms}$ depend on the ground-state magnetic density ${\rho}_{\mathrm{opt}}^{\rm mag}(\mathbf{r})$, i.e. they are equilibrium. Therefore, one cannot generate a training set containing non-equilibrium magnetic moments, which are required for mMTP fitting. Moreover, it is not possible to conserve the magnetic moments close to the selected target values $\mathrm{m}_{i}^{\mathrm{target}}$ chosen during active learning.

In this study, we used soft-constrained spin-polarized DFT calculations to obtain the resulting non-equilibrium magnetic moments $\mathrm{m}_i=\int_{\Omega_{i}} \rho^{\mathrm{mag}}(\mathbf{r}) d\mathbf{r}$ close to the target magnetic moments $\mathrm{m}_i^{\rm target}$. The underlying concept of this scheme is the following:

\begin{itemize}
    \item[1.] Minimize a functional:
    \begin{equation} \label{eq_main:dft_en}
    E^{\text{DFT}} = \min_{\rho} \left(E[\rho; \mathbf{R}, Z, L]+\lambda \sum_{i=1}^{N_{\rm atoms}}(\mathrm{m}_{i} - \mathrm{m}_i^{\rm target})^2 \right),
\end{equation}
where $\lambda$ is the chosen penalty weight.
    \item[2.] Introduce one or several criteria to stop calculations, e.g.
    \begin{equation} \label{eq_main:max_diff_magmom}
        \max_{i} \left|\mathrm{m}_{i} - \mathrm{m}_i^{\mathrm{target}}\right| < \Delta M_{\rm max},
    \end{equation}
    where $\Delta M_{\rm max}$ is the maximum difference between the absolute values of the target and the calculated magnetic moments. If Eq.~\eqref{eq_main:max_diff_magmom} is not valid, then increase $\lambda$ and minimize the functional in Eq.~\eqref{eq_main:dft_en} starting from the density and, therefore, the magnetic moments obtained in the previous step.
    \item[3.] Continue step~2 until the criterion for maximum difference in Eq.~\eqref{eq_main:max_diff_magmom} in magnetic moments is satisfied, i.e. magnetic moments are close to the target ones.
\end{itemize}

After the above procedure, we obtain magnetic moments close to the target ones and the DFT energies, forces, and stresses corresponding to these magnetic moments. This procedure enables us to calculate the configurations selected during any atomistic simulation using the active learning (AL) algorithm at the fixed atomic positions, atomic types, lattice vectors, and magnetic moments. 

Unlike in the paper~\cite{kotykhov2025_CrN} where we used hard constraint DFT (cDFT) calculations implemented in the ABINIT code \cite{gonze2020abinit, verstraete2025abinit}, here we used soft-constrained DFT calculations implemented in VASP. As we demonstrate below on the example of Fe--Pd, even soft-constrained spin-polarized calculations are sufficient for automated construction of a training set for fitting of magnetic MTP.

We note that the primary goal of soft-constrained DFT calculations is to constrain non-equilibrium magnetic moments $\mathrm{m}_{i}$. However, in VASP it is only possible to constrain the so-called smoothed magnetic moments. As a result, additional computational techniques are required to link the smoothed and target moments. For further details and a complete description of the methodology implemented, see subsection~\ref{sec:soft_constrained_scheme} in Supplementary Information. 

\subsection{Workflow for mMTP fitting on soft-constrained DFT calculations}

\begin{figure}[H]
\begin{center}
\includegraphics[width=0.99\columnwidth]{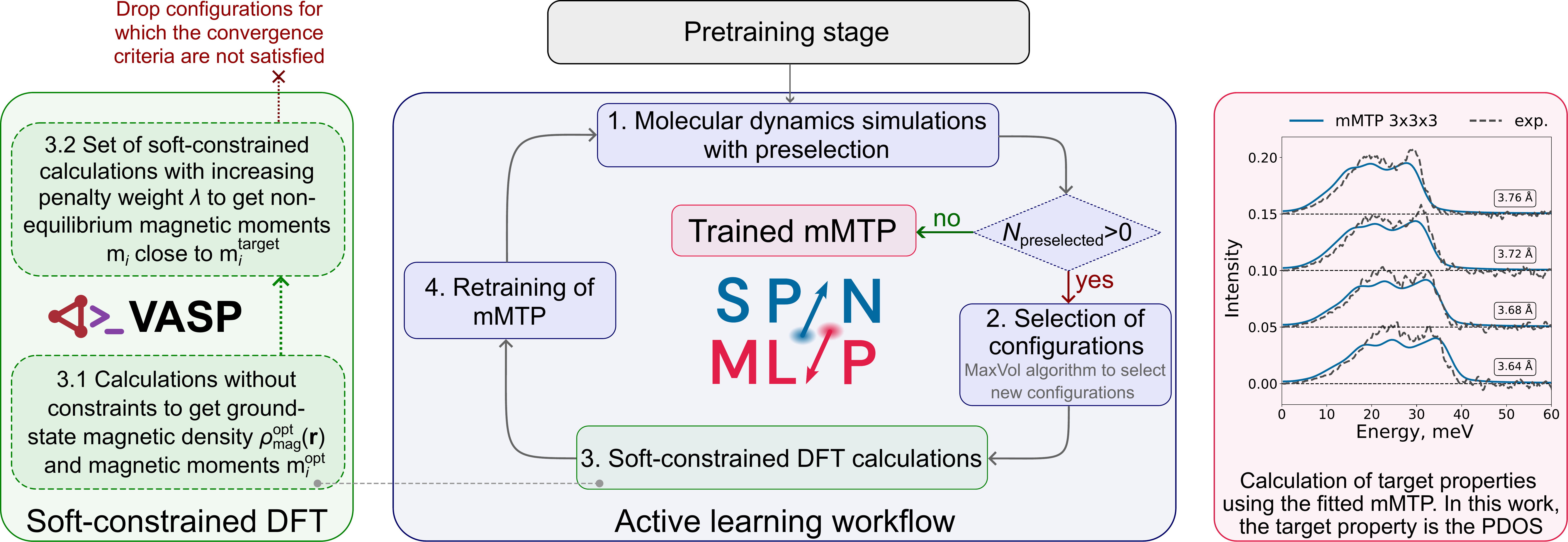}\caption{
Workflow of mMTP training incorporating soft‑constrained DFT calculations. The upper gray panel shows the pretraining stage, in which an initial mMTP is trained on DFT data, such as DFT MD trajectories. The central blue panel illustrates the active‑learning loop: (1) MD simulations proceed until no configuration is preselected; (2) otherwise, selection of configurations from the pool of preselected configurations is conducted; (3) soft‑constrained DFT calculations are performed for the selected configurations; (4) further retraining of mMTP is carried out on the updated training set. The left green panel details the soft‑constrained workflow: (3.1)~DFT calculations without constraints to obtain the ground‑state magnetic density $\rho_{\mathrm{opt}}^{\mathrm{mag}}$ and magnetic moments $\mathrm{m}_{i}^{\mathrm{opt}}$; (3.2)~a set of soft‑constrained DFT calculations with increasing penalty weight $\lambda$ to get non‑equilibrium magnetic moments $\mathrm{m}_{i}$ close to the target ones $\mathrm{m}_{i}^{\mathrm{target}}$. The right red panel illustrates the subsequent property calculations performed using the fitted mMTP, which in this work corresponds to the projected density of states (PDOS) of phonons.
}
\label{fig:scheme_main_panel}
\end{center}
\end{figure}

The general scheme of the fitting algorithm of mMTP on soft-constrained DFT calculations is shown in Fig.~\ref{fig:scheme_main_panel}. Before active learning commences, the potential is pretrained. Configurations derived from DFT molecular dynamics are utilized as an initial training set. After this, the active learning cycle itself is initiated, wherein the pretrained potential is used in a simulation (here, during molecular dynamics (MD) simulations) with preselection (extrapolation control) of configurations on the basis of extrapolation grade in Eq.~\eqref{eq:extrapolation_grade}. If at least one configuration has been preselected, then the selection of the most diverse configurations from the preselected pool is conducted. After that, DFT calculations for the selected configurations are performed in several stages. At the beginning, DFT calculations are carried out without constraints in order to determine the equilibrium magnetic moments and initial magnetic density for the next stage of DFT calculations. Next, a set of soft-constrained DFT calculations is carried out with a gradual increase in the penalty weight $\lambda$ and, in the case of convergence, the resulting magnetic moments $\mathrm{m}_i$ are close to the target ones $\mathrm{m}_i^{\rm target}$. After completion of the soft-constrained DFT calculations, the converged configurations are added to the training set, the potential is further retrained, after which the updated potential is used in modeling with preselection. The active learning cycle continues until no configuration is preselected. After that, the trained potential can be used to calculate the desired properties of materials.

\section{Results and discussion}

\subsection{Computational details}

\begin{figure}[!htb]
\begin{center}
\includegraphics[scale=0.1]{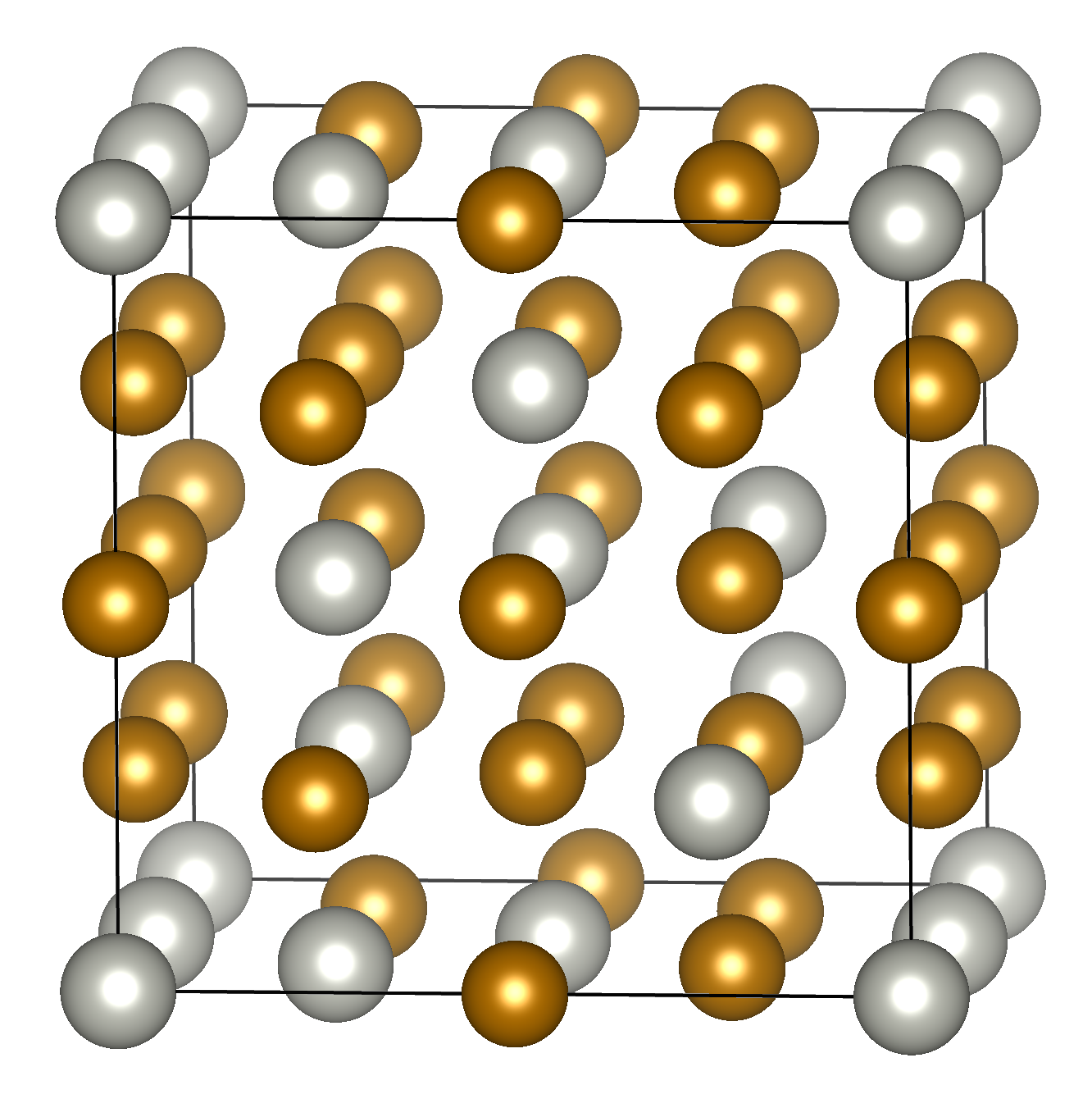}
\caption{The 32-atom fcc supercell of the Fe–Pd system, constructed from the conventional fcc cell with a lattice parameter of 3.72 ~\AA.}\label{fig:system_image}
\end{center}
\end{figure}

An initial training set was created with MD simulations using VASP for a face-centered cubic (fcc) Fe--Pd supercell containing 22 Fe atoms and 10 Pd atoms. The atomic species were randomly distributed over the lattice sites to approximate an $\text{Fe}_{0.68}\text{Pd}_{0.32}$ alloy composition. An example of structure is shown in Fig.~\ref{fig:system_image}. Simulations were performed for eight different lattice parameters (3.88, 3.84, 3.80, 3.76, 3.72, 3.68, 3.64, and 3.60~$\AA$) in the NVT ensemble at 300 K with time step of 1~femtosecond (fs) using the Nos\'e--Hoover thermostat with Nos\'e-mass corresponding to period of 40 time steps. These eight trajectories (one for each lattice parameter) lasted from 2.5 to 6.8~picoseconds (ps). We then took every fifth configuration from each trajectory without excluding configurations during thermalization and constructed an initial training set of 5823 configurations with equilibrium magnetic moments. For the subsequent property calculations performed with DFT, the first 500 steps of each trajectory were excluded, as these configurations corresponded to the thermalization period.

A test set comprising 320 configurations was also prepared. The first half of the test set consists of configurations with perturbed atomic positions and equilibrium magnetic moments, specifically, 20 atomic perturbations for each lattice parameter. Random atomic displacements were generated within the range $[-\sigma, \sigma]$, where the standard deviation $\sigma$ for each lattice parameter was predetermined from the training set. Subsequently, the magnetic moments of the resulting configurations were perturbed by up to ±5\% relative to their equilibrium values, yielding an additional 160 configurations with both perturbed atomic positions and perturbed magnetic moments.

We fitted mMTP of the 12-th level with $N_{\psi}$ = 2 magnetic and ${N_{\phi}}$ = 8 radial basis functions as defined in Eq.~\eqref{eq:radial_part}. The model has 415 trainable parameters and a cutoff radius $R_{\text{cut}}$ of 4~\AA. Weights of the loss function Eq.~\eqref{eq:loss} were set $w_e$ = 1.0, $w_f$ = 0.01 $\AA^2$, and $w_s$ = 0.001, analogously to the values chosen in the previous work \cite{Kotykhov2023-cDFT-mMTP}.

Active learning was performed during molecular dynamics simulations in the NVT ensemble at 300 K using a Nos\'e--Hoover thermostat with a damping parameter of 3 ps. The simulation time step was set to 1 fs. For each lattice parameter, a single trajectory was initiated and propagated until either the $\gamma_{\text{up}}$ threshold was exceeded or the maximum iteration limit of 50000 was reached.
The ferromagnetic state was used to set the initial magnetic moments for mMTP, with values of 2.65 $\mu_{B}$ for Fe atoms and 0.3 $\mu_{B}$ for Pd atoms. Within each active learning iteration, configurations were preselected from several trajectories at different volumes. This leads to a substantial increase in the number of candidates and lengthens the selection process of preselected configurations. To avoid this, here stricter thresholds of $\gamma_{\text{low}} = 3$ and $\gamma_{\text{up}} = 5$ were used instead of the typical thresholds of 2 and 10 utilized in most of the previous papers (see, e.g.~\cite{novikov2018-rpmd-al-mtp}). Details on the recommended values of the thresholds $\gamma_{\text{low}}$ and $\gamma_{\text{up}}$ can be found in Refs. \cite{podryabinkin2017active, novikov2021mlip}. Each MD time step is preceded by a search for equilibrium magnetic moments. Magnetic moment optimization (equilibration) is completed when the norms of all magnetic forces do not exceed 0.001~eV per Bohr magneton. Thus, preselection of configurations occurs both during the equilibration of magnetic moments and during changes in atomic positions during molecular dynamics. This approach ensures that the training set includes enough diversity for both magnetic and spatial degrees of freedom.

To calculate configurations for a training set, we conducted DFT calculations using the projector augmented plane wave method, with the Vienna ab initio Simulation Package (VASP)~\cite{Kresse:1996,Kresse:1999}. All the calculations were performed within the generalized gradient approximation (GGA) in the Perdew--Burke--Ernzerhof (PBE) form~\cite{Perdew:PRL:1996, Perdew:PRL:1997} for the exchange-correlation (XC) functional. The electron--ion interaction was described with the projector augmented wave (PAW) method~\cite{Blochl:1994}, using the potentials of version~54. The plane-wave kinetic energy cut-off ($E_{\mathrm{cut}}$) was 500~eV. The $k$-mesh in the reciprocal space was $\Gamma$-centered with $k$-spacing less than 0.2~\AA$^{-1}$ for all crystal lattices. The convergence criterion for energies in the non-constrained calculations was $10^{-6}$~eV. Brillouin-zone integration was performed using Methfessel--Paxton smearing of first order with a smearing width of 0.1~eV. Further details of DFT/soft-constrained DFT calculations are described in Supplementary Information in section~\ref{sec:soft_constrained_dft}. 

\begin{figure}[!htb]
\begin{center}
    \includegraphics[scale=0.25]{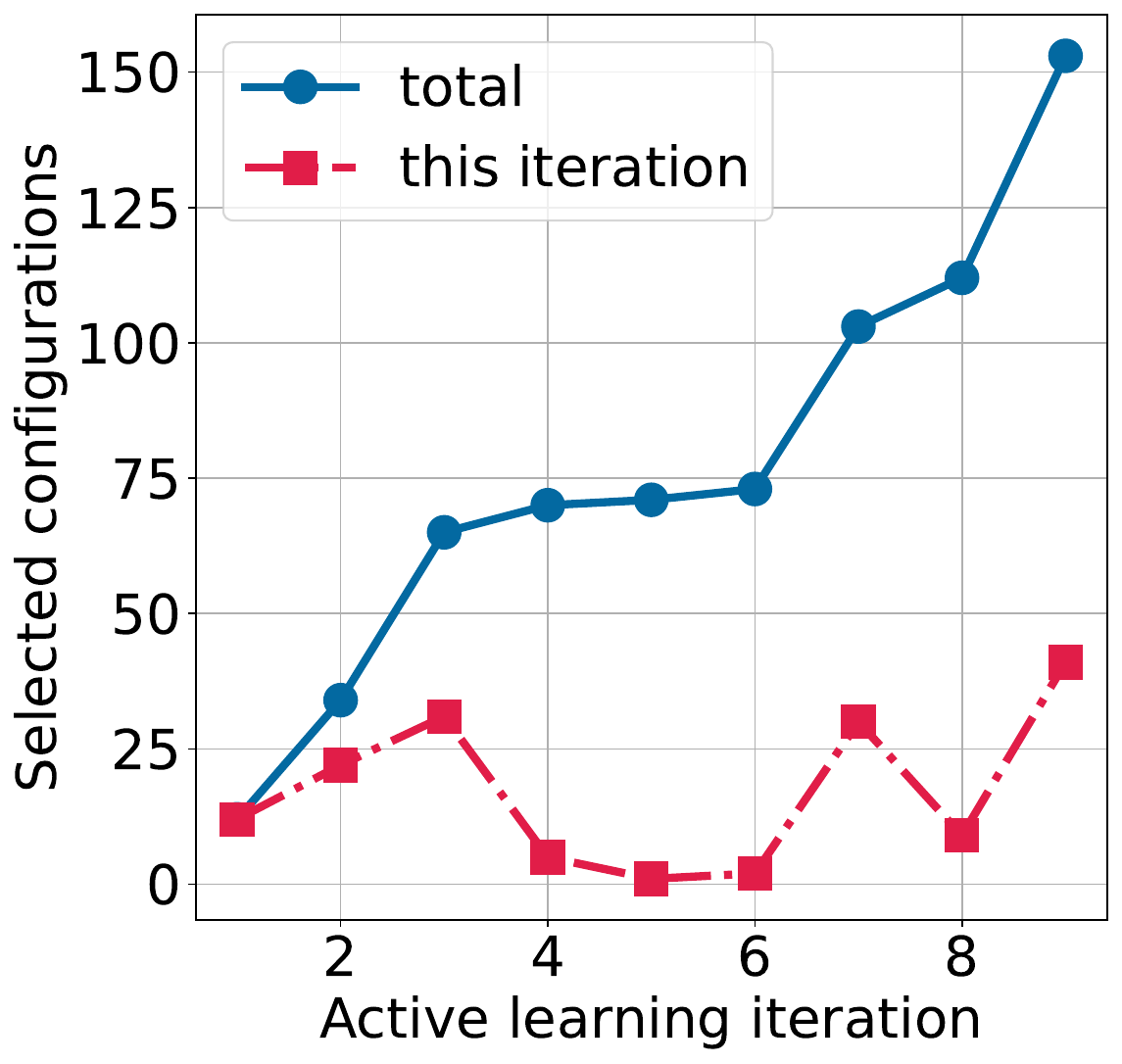}
\caption{The number of selected configurations during active learning in total and at each iteration.}\label{fig:N_selected}
\end{center}
\end{figure}

The active learning procedure consisted of 9 iterations, resulting in the addition of 153 configurations to the training set. The distribution of selected configurations across active learning iterations is shown in Fig.~\ref{fig:N_selected}.

After the active learning stage, NVT molecular dynamics simulations were performed using the same simulation settings as those employed during active learning to compute the velocity autocorrelators for the subsequent calculation of the phonon DOS. For each lattice parameter within the range of 3.60 to 3.88~\AA~with a step of 0.04~\AA, 16 independent trajectories of 50000 MD steps each were generated at 300 K. The first 6000 steps of each trajectory were discarded for thermalization and were therefore excluded from the subsequent calculation of the properties.

\subsection{Fitting and test errors}

A comparison of the training and test errors, as well as the magnetic moment equilibration convergence, for each active learning iteration for the mMTP potential is presented in Table~\ref{tab:fitting_errors}. Here, equilibration convergence is defined as the fraction of initial configurations, for each of the eight lattice parameters, that reach a gradient magnitude less than 1 meV/$\mu_B$ within 1500 energy minimization steps. After active learning, the root-mean-square errors (RMSEs) evaluated on the initial training set decrease by 16\%, 1\%, and 5\% for energies, forces, and stresses, respectively. On the test set, the energy and stress RMSEs decrease substantially, by 39\% and 39\%, respectively, while the force RMSE increases only slightly, by 4\%. In addition, the convergence rate of the magnetic moment equilibration reaches 100\% after active learning. This complete equilibration convergence is important for ensuring the reliability of subsequent molecular dynamics simulations performed with the trained potential.

\begin{table}[!htb]
\centering
\caption{Root-mean-square errors (RMSEs) for energies (meV/atom), forces (meV/\AA), and stresses (GPa) on the initial training set, current training set (with selected configurations) and test set, together with magnetic-moment equilibration convergence, before active learning, for each active learning iteration, and after active learning. Convergence is defined as the fraction of successful magnetic moment equilibrations for eight initial configurations with different lattice parameters.}
\label{tab:fitting_errors}
\begin{tabular}{c c c c c c c c c c c}
\toprule
\multirow{2}{*}{Potentials} & \multirow{2}{*}{\parbox{2.5cm}{\centering Convergence (\%)}} & \multicolumn{3}{c}{Initial training set} & \multicolumn{3}{c}{Current training set} & \multicolumn{3}{c}{Test set} \\
\cmidrule(lr){3-5} \cmidrule(lr){6-8} \cmidrule(lr){9-11}
                            & & E & F & S & E & F & S & E & F & S \\
\midrule
Before AL                   & 62.5                         & 1.61 & 109.5 & 0.44 & 1.61 & 109.5 & 0.44 & 2.82 & 117.1 & 0.95 \\
AL 1                        & 75                           & 1.41 & 108.8 & 0.44 & 1.46 & 108.9 & 0.45 & 5.24 & 139.7 & 0.78 \\
AL 2                        & 100                          & 1.42 & 109.0 & 0.44 & 2.55 & 109.2 & 0.45 & 4.29 & 117.6 & 0.82 \\
AL 3                        & 50                           & 1.34 & 108.9 & 0.44 & 1.73 & 109.2 & 0.45 & 2.98 & 127.3 & 0.71 \\
AL 4                        & 100                          & 1.32 & 107.9 & 0.43 & 1.75 & 108.3 & 0.44 & 2.98 & 133.0 & 0.71 \\
AL 5                        & 100                          & 1.32 & 108.1 & 0.42 & 1.68 & 108.5 & 0.43 & 2.99 & 131.7 & 0.73 \\
AL 6                        & 100                          & 1.35 & 108.0 & 0.42 & 1.69 & 108.4 & 0.43 & 3.63 & 125.0 & 0.71 \\
AL 7                        & 75                           & 1.39 & 108.0 & 0.42 & 1.84 & 108.7 & 0.43 & 2.48 & 141.7 & 0.59 \\
AL 8                        & 100                          & 1.37 & 108.1 & 0.42 & 1.83 & 108.8 & 0.43 & 1.71 & 133.8 & 0.59 \\
After AL                    & 100                          & 1.36 & 108.0 & 0.42 & 1.86 & 109.0 & 0.43 & 1.72 & 121.4 & 0.58 \\
\bottomrule
\end{tabular}
\end{table}

\subsection{Magnetization}

\begin{figure}[!htb]
\begin{center}
\includegraphics[scale=0.3]{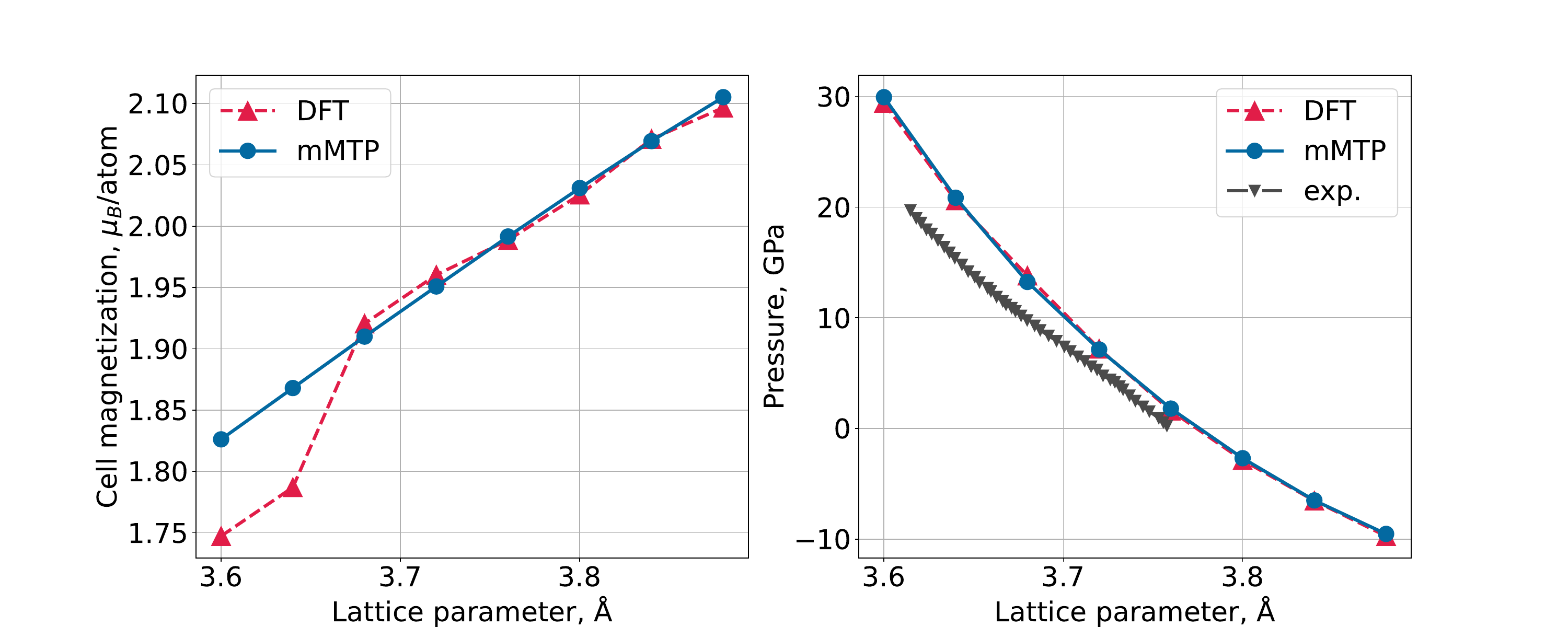}\caption{Dependence of magnetization and pressure on lattice parameter at $T=300$~K. The experimental data were taken from~\cite{priesen2025thermodynamic}.}\label{fig:magnetization_and_press}
\end{center}
\end{figure}

Experimental findings for $\text{Fe}_{0.68}\text{Pd}_{0.32}$ reveal a pressure-induced reduction in magnetization of this crystal with increasing pressure~\cite{matsushita2002anomalous, priesen2025thermodynamic}. The magnetization and pressure as functions of the lattice parameter are shown in Fig.~\ref{fig:magnetization_and_press}, alongside experimental data from~\cite{priesen2025thermodynamic}. The values of magnetization and pressure were obtained by averaging over iterations in molecular dynamics after thermalization for a supercell containing 32 atoms. Both DFT and trained mMTP qualitatively reproduce the experimentally observed decrease in magnetization with increasing pressure (decreasing lattice parameter). The results for magnetization and pressure dependence obtained using mMTP are in good agreement with DFT-calculated values, except for the magnetization at the two smallest lattice parameters. Since the ferromagnetic state was chosen to define the initial magnetic moments, and equilibrating magnetic moments using mMTP does not result in spin flips, unlike DFT, this leads to an overestimation of the magnetization. For mMTP, MD simulations including Monte Carlo (MC) spin flips (MDMC) could be used to find the global energy minimum; however, such an approach is beyond the scope of this study.

\subsection{Density of states}

\begin{figure}[!htb]
\begin{center}
\includegraphics[scale=0.4]{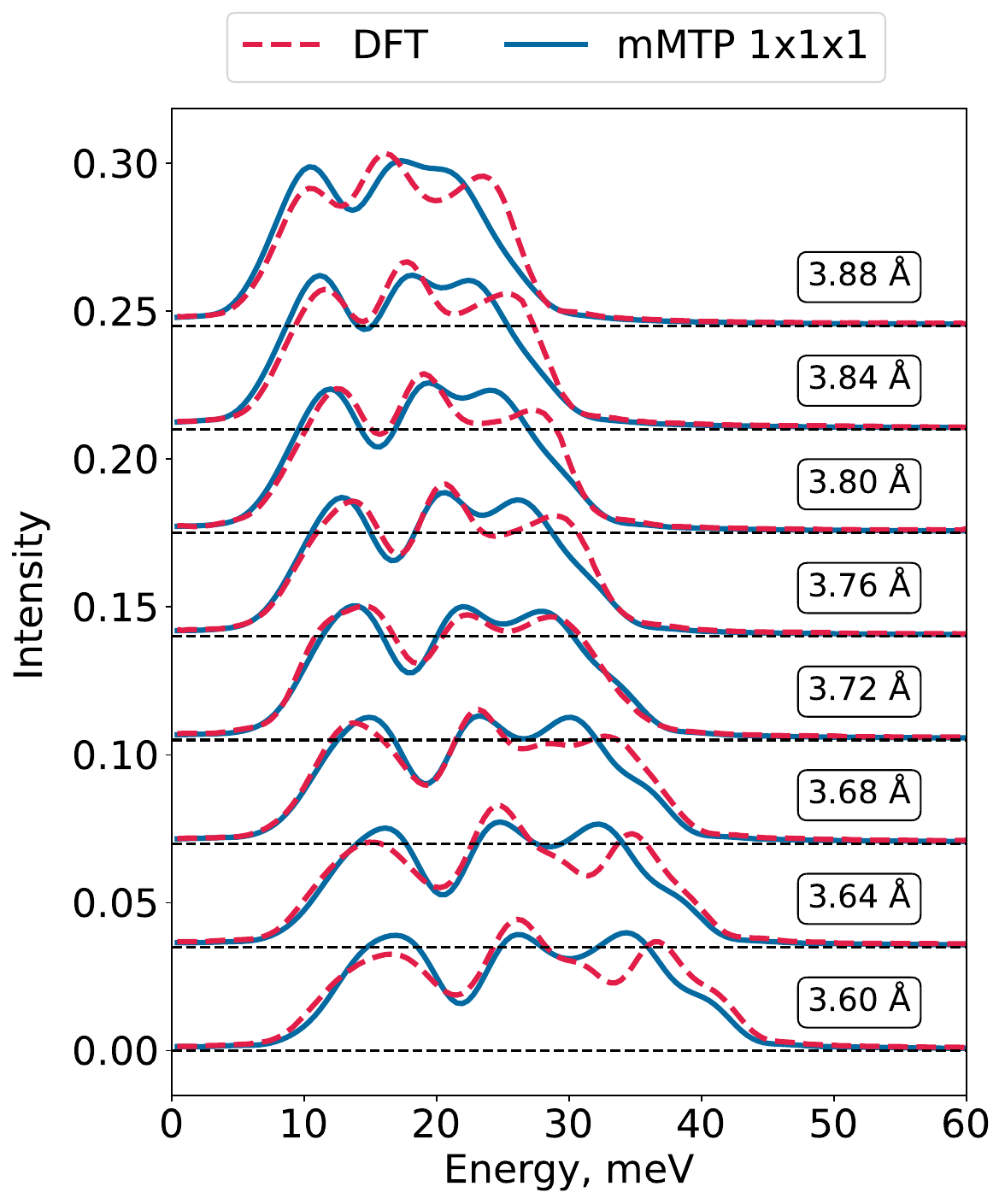}\caption{Total phonon DOS for $\text{Fe}_{0.68}\text{Pd}_{0.32}$ calculated with mMTP and DFT for different lattice parameters at 300~K.}\label{fig:total_DOS_MTP_vs_DFT}
\end{center}
\end{figure}

For eight lattice parameters in the range from 3.60 to 3.88~\AA, molecular dynamics was performed in the NVT ensemble at a temperature of 300~K using different numbers of atoms in the Fe--Pd supercell. Phonon density of states (DOS) was computed by taking the Fourier transform of the velocity autocorrelation function. A comparison of the phonon DOS obtained from mMTP and DFT for non-replicated cell (i.e., 32~atoms) is shown in Fig.~\ref{fig:total_DOS_MTP_vs_DFT}. The DOS obtained from mMTP agrees qualitatively with the DFT benchmark across most frequencies.

\begin{figure}[!htb]
\begin{center}
\includegraphics[scale=0.4]{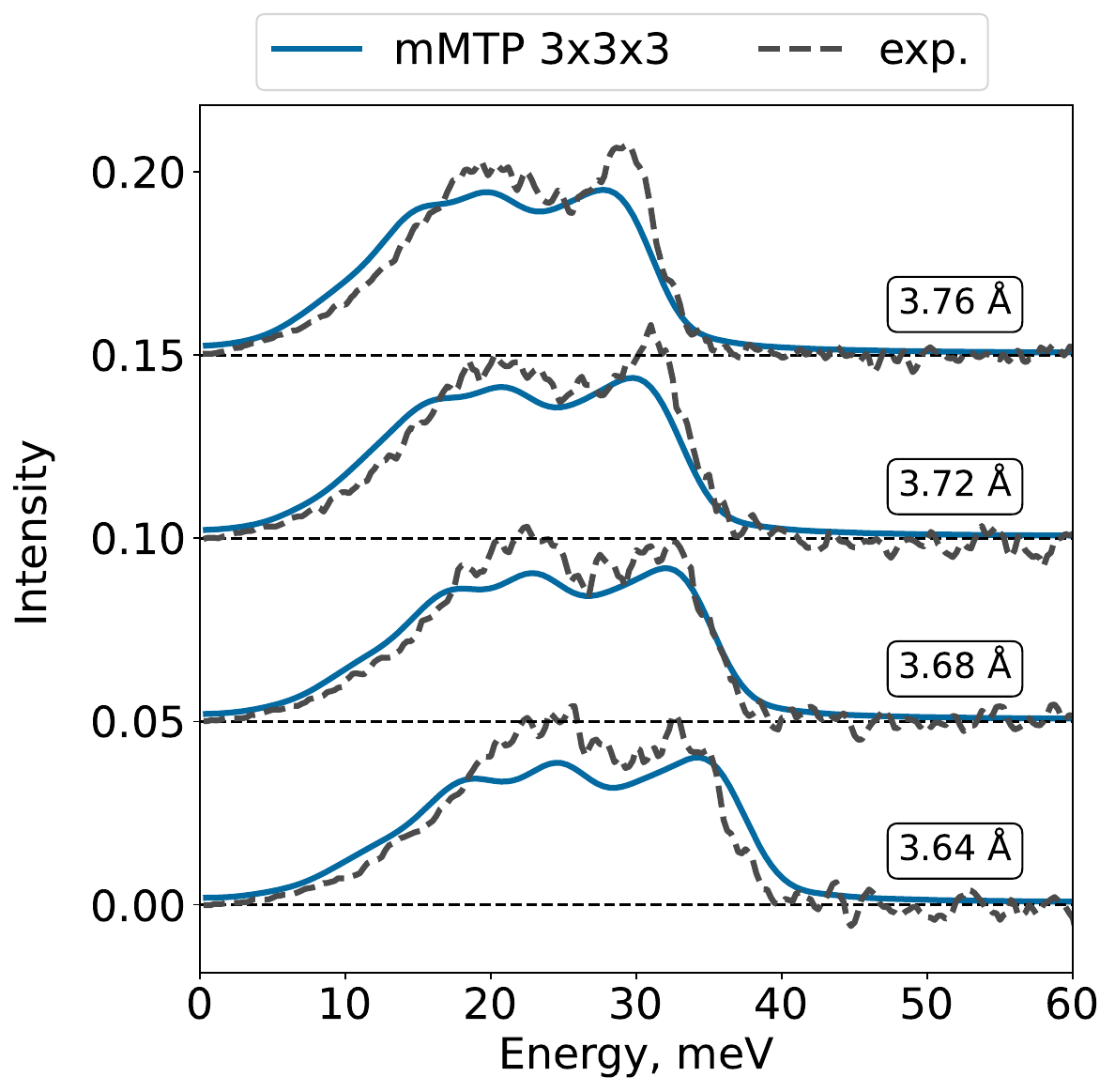}\caption{Fe phonon DOS for $\text{Fe}_{0.68}\text{Pd}_{0.32}$ calculated with mMTP and compared with experiment from~\cite{priesen2025thermodynamic} for different corresponding lattice parameters at 300~K.}\label{fig:Fe_DOS_MTP_vs_exp}
\end{center}
\end{figure}

We also present the Fe phonon density of states for $\text{Fe}_{0.68}\text{Pd}_{0.32}$ in Fig.~\ref{fig:Fe_DOS_MTP_vs_exp}, comparing it with the experimental results from work~\cite{priesen2025thermodynamic}. Simulations with mMTP were carried out with a replicated $3\times3\times3$ cell comprising 864~atoms and lattice parameters were chosen close to the experimental values. We observe that mMTP correctly captures the qualitative evolution of the DOS with pressure, namely its shift toward higher frequencies as the lattice parameter decreases.

\section{Conclusion}

In this work, we presented the workflow for active learning of magnetic Moment Tensor Potential (mMTP) during molecular dynamics simulations on the data obtained with density functional theory (DFT) and soft-constrained DFT calculations. Active learning of mMTP was implemented and conducted in the spin-MLIP package~\cite{spin-MLIP}, molecular dynamics (MD) simulations were carried out in LAMMPS~\cite{plimpton1995fast,thompson2022lammps}, and DFT calculations were performed in the VASP package~\cite{Kresse:1996,Kresse:1999}. 

The proposed methodology was tested for active training of mMTP on $\text{Fe}_{0.68}\text{Pd}_{0.32}$. A training set of 5976 configurations including 153 configurations with non-equilibrium magnetic moments which were calculated with soft-constrained DFT was prepared. The training errors for energy, forces, and stresses after active learning are 1.86~meV/atom, 109~meV/\AA, and 0.43~GPa, respectively. We used this actively trained potential to model the phonon density of states (DOS) and cell magnetization at lattice parameters from 3.60 to 3.88~\AA with NVT molecular dynamics at 300~K. The cell magnetization obtained using mMTP is in excellent agreement with the DFT values except for the lattice parameters of 3.60 and 3.64~\AA. A possible reason for this discrepancy is that the equilibration of magnetic moments starting from the ferromagnetic state using mMTP does not result in spin flips, unlike DFT. This leads to an overestimation of the magnetization with mMTP for compressed cells. The phonon DOSs calculated with mMTP are consistent with the experimental data. We observed that mMTP correctly captures the shift of DOSs toward higher frequencies as the lattice parameter decreases (or, pressure increases).

In the future, the described workflow for active learning together with soft-constrained DFT calculations can be generalized to non-collinear magnetic potentials with configuration selection being performed during spin-lattice dynamics. Another possible direction is an implementation of molecular dynamics (MD) simulations including Monte Carlo (MC) spin flips (MDMC) which enables finding deeper energy minima and achieving results that better agree with the DFT ones.

\section{Acknowledgments}

The publication was prepared within the framework of the Academic Fund Program at HSE University (grant No.~26-00-054 ``Machine learning in atomistic modeling''). This research was supported in part by computational resources of HPC facilities at the HSE University~\cite{kostenetskiy2021hpc}.

\section{Supplementary information}

Details of soft-constrained spin-polarized DFT calculations are given in supplementary information.

\section{Conflict of interest}

The authors have no conflict of interest to disclose.

\section{Data availability}

The spin-MLIP package is available at~\cite{spin-MLIP} and its interface with the LAMMPS code is available at~\cite{spin-MLIP-LAMMPS}. We also provide a tutorial available at the link~\cite{pipeline} with an example of active learning of mMTPs on DFT/soft-constrained DFT calculations for the Fe--Pd crystal. Finally, the training set as well as the fitted mMTP are available at the same link.

\beginsupplement
\section*{Supplementary Information: Active learning of collinear magnetic Moment Tensor Potentials using the spin-MLIP package from soft-constrained spin-polarized DFT calculations: a case study of Fe--Pd}\label{sec:si}

Our current workflow has been implemented and tested only for conserving collinear magnetic moments, in which all magnetic moments $\mathrm{m}_i$ are aligned along a single direction and thus have only one non-zero component, either $x$, or $y$, or $z$. Therefore, we use only scalar magnetic moments $\mathrm{m}_i$ instead of the vector $\mathbf{m}_i$. The requirement to carry out non-collinear calculations arises from limitations in VASP's implementation of soft-constrained spin-polarized calculations. However, with only minor modifications, our workflow for collinear soft-constrained calculations can be extended to the non-collinear case.

\section{Soft-constrained spin-polarized DFT calculations} \label{sec:soft_constrained_dft}

\subsection{VASP notations and types of magnetic moments}

\begin{table}[H]
\small
\begingroup
\renewcommand{\arraystretch}{1.2}
\centering
\caption{Notation for VASP files, INCAR tags, definitions of different magnetic moments, and other technical terms used in this work.}
\label{tab:notations}
\begin{tabular}{ll}
\toprule
Notation           & Description                                          \\ \midrule
\multicolumn{2}{c}{\cellcolor[HTML]{EFEFEF}VASP files}               \\ \midrule
INCAR         & File with input parameters for the VASP calculations                   \\
POTCAR        & Pseudopotential with data for all atomic species                   \\
OUTCAR        & Output file with detailed information about the VASP run                     \\
OSZICAR       & Convergence information, including $\mathrm{M}_{\mathrm{int}}$ ($\mathrm{m}_{i}^{\mathrm{WS}}$) and $\mathrm{MW}_{\mathrm{int}}$ ($\mathrm{m}_{i}^{\mathrm{sWS}}$) at each ionic step   \\
WAVECAR       & Wave function file                                     \\
CHGCAR        & Charge density file                                      \\ \midrule
\multicolumn{2}{c}{\cellcolor[HTML]{EFEFEF}INCAR tags}                     \\ \midrule
MAGMOM        & Initial magnetic moments assigned to each atom                \\
M\_CONSTR     & Desired local magnetic moments. The smoothed magnetic moments $\mathrm{MW}_{\mathrm{int}}$ ($\mathrm{m}_{i}^{\mathrm{sWS}}$) \\ & are close to them after the soft-constrained calculations \\
LNONCOLLINEAR & Enables non-collinear magnetic calculations      \\
I\_CONSTRAINED\_M  &  Switches on the constrained local moments approach  \\ 
LWAVE         & Specifies whether the wave function is written    \\
LCHARG        & Specifies whether the charge density is written    \\ 
RWIGS        &  Wigner-Seitz radius for each atomic type. Used for $\mathrm{M}_{\mathrm{int}}$ integration    \\ \midrule

\multicolumn{2}{c}{\cellcolor[HTML]{EFEFEF}Other notations}                     \\ \midrule
$\mathrm{M}_{\mathrm{int}}$        & Moment integrated over atom‑centered spheres with Wigner-Seitz radius ($\mathrm{m}_i^{\mathrm{WS}}$)      \\
$\mathrm{MW}_{\mathrm{int}}$       & Smoothed moments integrated over atom‑centered spheres of  Wigner-Seitz radius ($\mathrm{m}_i^{\mathrm{sWS}}$)                \\ \midrule

\multicolumn{2}{c}{\cellcolor[HTML]{EFEFEF}Types of magnetic moments}                     \\ \midrule
$\mathrm{m}_{i}$ & Magnetic moments from (PAW) region integration (M, located in OUTCAR)  \\
$\mathrm{m}_{i}^{\mathrm{WS}}$ & Magnetic moments from Wigner-Seitz sphere integration ($\mathrm{M}_{\mathrm{int}}$, located in OSZICAR) \\
$\mathrm{m}_{i}^{\mathrm{sWS}}$ & Smoothed magnetic moments from Wigner-Seitz sphere integration ($\mathrm{MW}_{\mathrm{int}}$, located in OSZICAR) \\
$\mathrm{m}_{i}^{\mathrm{target}}$ & Target magnetic moments (required to retrain mMTP after selection) \\
$\mathrm{m}_{i}^{\mathrm{constr}}$ & Imposed constraint magnitude on smoothed magnetic moments (M\_CONSTR, specified in INCAR) \\
$\widetilde{\mathrm{m}}_{i}^{\mathrm{refined}}$ & Refined constraint magnitude on smoothed magnetic moments used to achieve $\mathrm{m}_{i} \simeq \mathrm{m}_{i}^{\mathrm{target}}$ \\ & after the last stage of soft-constrained DFT calculations (M\_CONSTR, specified in INCAR) \\ \bottomrule

\end{tabular}
\endgroup
\end{table}

\subsection{Definitions of VASP magnetic moments}\label{sec:vasp_mag_moments}

To formulate an algorithm, we introduce the so-called smoothed magnetic moments, which are the only option to constrain in VASP, using the M\_CONSTR tag: 
\begin{equation} \label{eq:smooth_magmom}
{\rm m}_{i}^{\mathrm{sWS}} = \int_{\Omega_{i}^{\mathrm{WS}}} \rho_{\mathrm{mag}}(\mathbf{r}) F_{i}(|\mathbf{r}|) \, d\mathbf{r},
\end{equation}
where $\Omega_i^{\mathrm{WS}}$ is the atom‑centered sphere of the radius given by the RWIGS tag, and $F_i(|\mathbf{r}|)$ is a function that equals unity near the center of the sphere and smoothly decays to zero at the boundary.

The corresponding unsmoothed magnetic moments (denoted as $\mathrm{M}_{\mathrm{int}}$ in OSZICAR) are computed as
\begin{equation} \label{eq:nonsmooth_magmom}
\mathrm{m}_i^{\mathrm{WS}} = \int_{\Omega_i^{\mathrm{WS}}} \rho_{\mathrm{mag}}(\mathbf{r}) \, \mathrm{d}\mathbf{r}.
\end{equation}

\begin{figure}[H]
\begin{center}
\includegraphics[width=0.99\columnwidth]{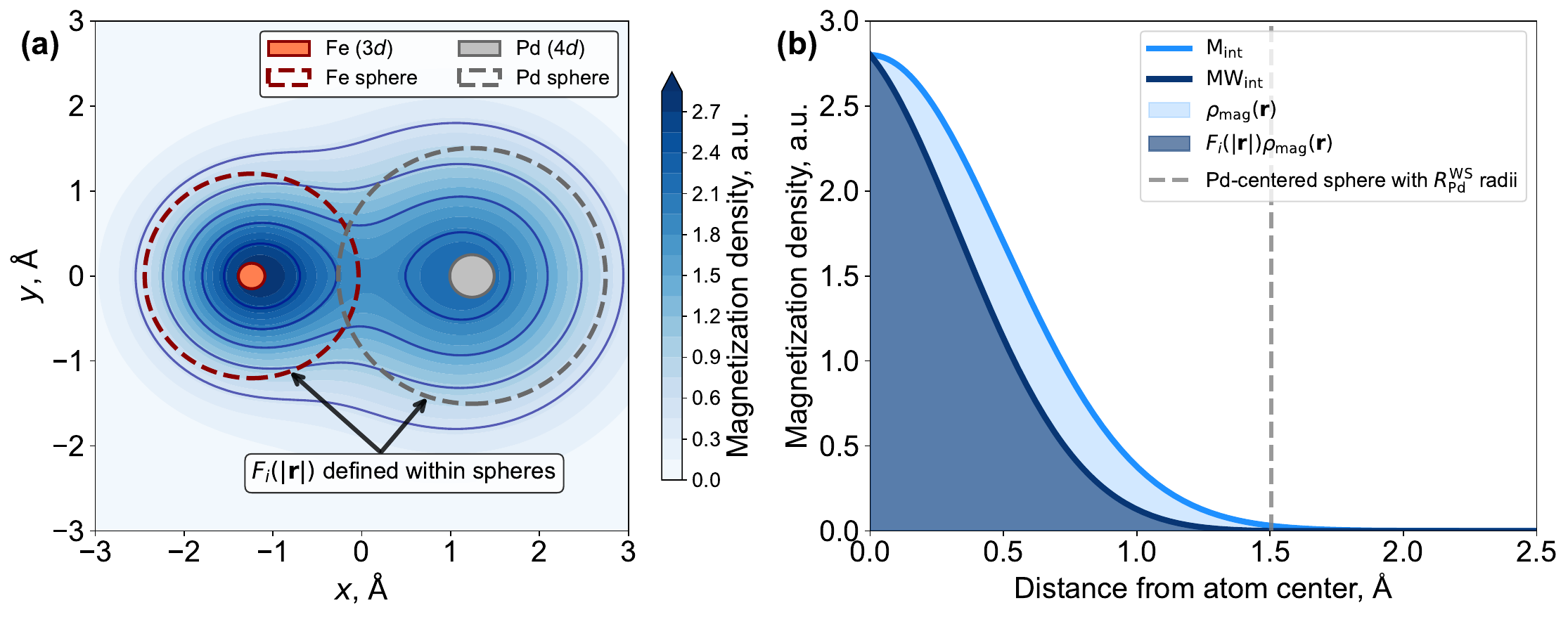}\caption{
Magnetic density around Fe--Pd atoms and the effect of the smoothing function.  
(a) Two-dimensional magnetic density $\rho_\mathrm{mag}(\mathbf{r})$ (blue) around an Fe atom (red) and a Pd atom (silver), with the projections of the atom-centered spheres onto the $xy$ plane outlined by dashed circles corresponding to the Wigner-Seitz radii, as specified by the RWIGS tag in VASP.  
(b) Illustration of the origin of the difference between $\mathrm{M}_{\mathrm{int}}$ and $\mathrm{MW}_{\mathrm{int}}$ for a Pd atom. Radial magnetic profiles: the raw $\rho_\mathrm{mag}(\mathbf{r})$ (sky blue) versus the smoothed product $F_i(|\mathbf{r}|)\,\rho_\mathrm{mag}(\mathbf{r})$ (dark blue). The function $F_i(|\mathbf{r}|)$ ramps from unity at the atomic core to zero at the boundary of the atom-centered sphere. The gray dashed line indicates the Wigner-Seitz radius of the Pd atom ($R_{\mathrm{Pd}}^{\mathrm{WS}}$), up to which the integration of the magnetic density is performed.
}\label{fig:smoothing_method}
\end{center}
\end{figure}

The difference between $\mathrm{m}_{i}^{\mathrm{WS}}$ ($\mathrm{M}_{\mathrm{int}}$) and $\mathrm{m}_{i}^{\mathrm{sWS}}$ ($\mathrm{MW}_{\mathrm{int}}$) is illustrated in Fig.~\ref{fig:smoothing_method}, using arbitrary units for clarity. Both $\mathrm{m}_{i}^{\mathrm{WS}}$ and $\mathrm{m}_{i}^{\mathrm{sWS}}$ are written in the OSZICAR file.

There is also a third type of magnetic moments written in OUTCAR file. These moments are integrated over the projector-augmented wave (PAW) regions and obtained by decomposing the Kohn-Sham orbitals into angular-momentum ($l$) and magnetic quantum number ($m$) components within the PAW spheres~\cite{Blochl:1994}. In this case, Eq.~\eqref{eq:nonsmooth_magmom} is rewritten as
\begin{equation} \label{eq:outcar_magmom}
\mathrm{m}_i = \int_{\Omega_i^{\mathrm{PAW}}} \rho_{\mathrm{mag}}(\mathbf{r}) \, \mathrm{d}\mathbf{r},
\end{equation}
where $\Omega_i^{\mathrm{PAW}}$ denotes the PAW augmentation region around atom $i$ and $\rho_{\mathrm{mag}}(\mathbf{r})$ is the magnetic density. The aim of our algorithm of soft-constrained DFT calculations is to obtain $\mathrm{m}_i \simeq \mathrm{m}_i^{\mathrm{target}}$.

\subsection{Scheme and technical aspects of soft-constrained workflow}\label{sec:soft_constrained_scheme}

\begin{figure}[H]
\begin{center}
\includegraphics[width=0.99\columnwidth]{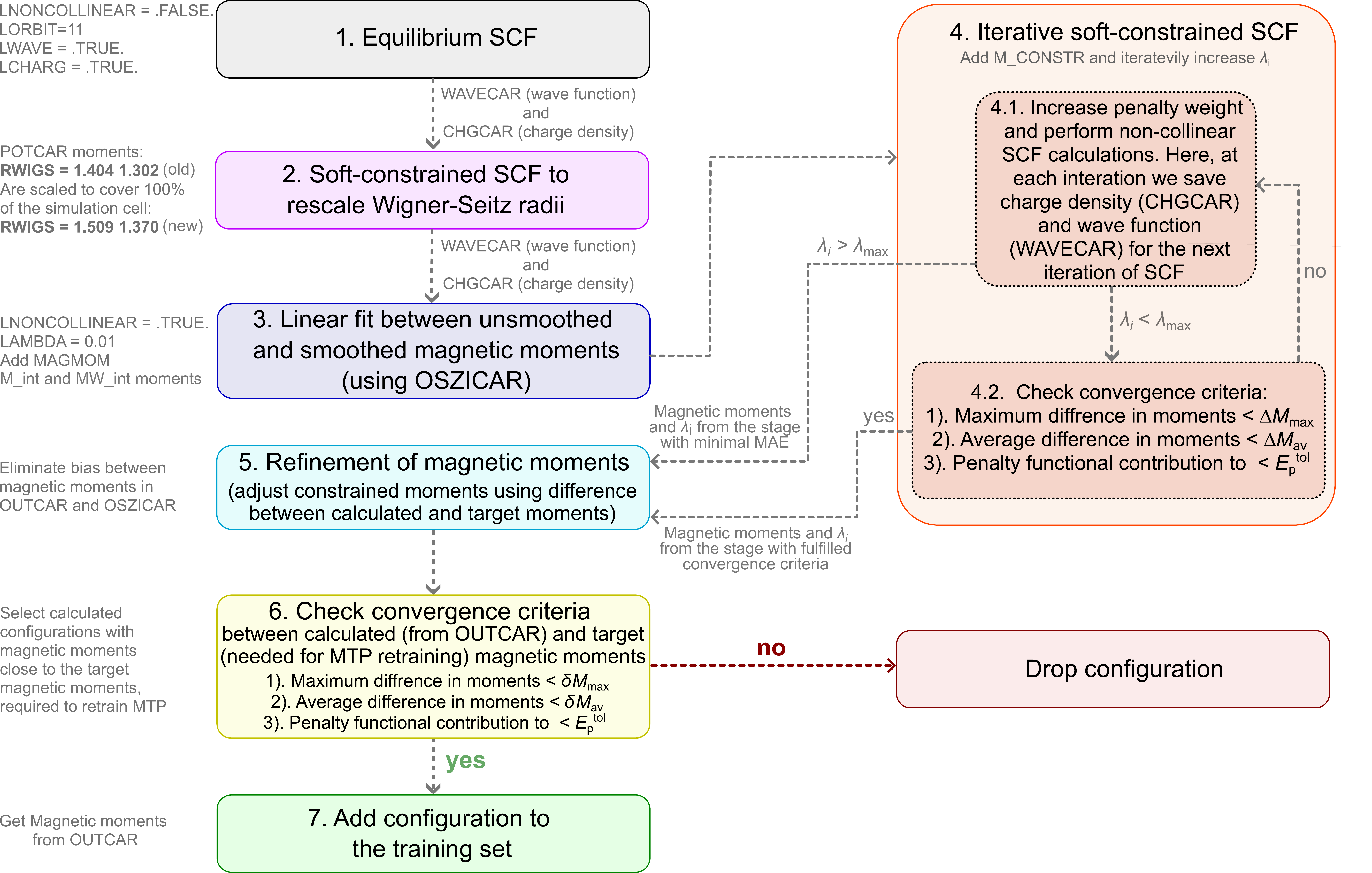}\caption{
Scheme of soft-constrained DFT calculations. (1) Equilibrium self-consistent field (SCF) calculation. (2) Soft-constrained SCF calculation in which the Wigner-Seitz radii of the atom-centered spheres are rescaled proportionally from their initial values in POTCAR to cover approximately 100~\% of the simulation cell. (3) Linear fit is performed between the unsmoothed magnetic moments $\mathrm{m}_{i}^{\mathrm{WS}}$ ($\mathrm{M}_{\mathrm{int}}$) and the smoothed magnetic moments $\mathrm{m}_{i}^{\mathrm{sWS}}$ ($\mathrm{MW}_{\mathrm{int}}$). (4) Iterative soft-constrained SCF calculations: the penalty weight $\lambda$ is increased until the maximum and average magnetic moment differences decrease below $\Delta M_{\mathrm{max}}$ and $\Delta M_{\mathrm{av}}$, respectively, and the contribution of the penalty functional to the total energy is smaller than $E_p^{\mathrm{tol}}$ (see the criteria in Eqs.~\eqref{eq:max_diff_magmom},\,\eqref{eq:av_diff_magmom}, and~\eqref{eq:penalty_en}). If these criteria are satisfied, the configuration proceeds to the final refinement stage with the current $\lambda_i$. Otherwise, if the criteria are not satisfied at $\lambda_i = \lambda_{\mathrm{max}}$, the configuration calculated with the value of $\lambda_i$ that yields smoothed magnetic moments $\mathrm{m}_{i}^{\mathrm{sWS}}$ closest to the constraining smoothed magnetic moments $\mathrm{m}_{i}^{\mathrm{constr}}$ is sent for refinement. (5) At the refinement stage, the constrained magnetic moments in INCAR (via the M\_CONSTR tag) are adjusted using the difference between the magnetic moments in OSZICAR ($\mathrm{M}_{\mathrm{int}}$) and OUTCAR ($\mathrm{m}_{i}$). After this refinement step, the magnetic moments $\mathrm{m}_i$ reported in OUTCAR file are approximately equal to the target moments $\mathrm{m}_i^{\mathrm{target}}$. (6) Apply the same MaxAE, MAE, and penalty-energy thresholds as in Eqs.~\eqref{eq:max_diff_magmom},\,\eqref{eq:av_diff_magmom}, and~\eqref{eq:penalty_en}, but now to $\mathrm{m}_{i}$ versus $\mathrm{m}_{i}^{\mathrm{target}}$. Configurations that satisfy these criteria are added to the training set (panel~7); otherwise, they are discarded.
}
\label{fig:scheme_soft}
\end{center}
\end{figure}

From a conceptual point of view, the scheme of soft-constrained DFT calculations is defined in subsection~\ref{SoftConstrainedDFT_main} for $\mathrm{m}_{i}$ and $\mathrm{m}_{i}^{\rm target}$. However, as it was mentioned in the main text, in practice it is used for $\mathrm{m}_{i}^{\mathrm{sWS}}$ due to the specifics of the VASP package. The complete scheme of soft-constrained DFT calculations to get $\mathrm{m}_{i} \simeq \mathrm{m}_{i}^{\mathrm{target}}$ is presented in Fig.~\ref{fig:scheme_soft} and described in the text below.

\begin{itemize}
    \item[1.] At the beginning, equilibrium self-consistent field calculations (collinear, without constraints on magnetic moments) are performed to obtain an initial approximation of the wave function and ground-state charge density $\rho_{\mathrm{opt}}^{\mathrm{mag}}(\mathbf{r})$, thereby accelerating the subsequent non-collinear calculations.
    \item[2.] After this, non-collinear soft-constrained self-consistent field calculations are performed to rescale the Wigner-Seitz radii. This step is added because the moments $\mathrm{m}_i^{\mathrm{sWS}}$ and $\mathrm{m}_i^{\mathrm{WS}}$ reported in the OSZICAR files are obtained by integrating over atom-centered spheres with Wigner-Seitz radii (the RWIGS tag). The differences in the integration schemes between $\mathrm{m}_i^{\mathrm{WS}}$ and $\mathrm{m}_i$ lead to a small discrepancy between them. However, this difference can be reduced if the integration spheres cover almost 100~\% of the simulation cell by rescaling the Wigner-Seitz radii (more details can be found in subsection~\ref{sec:rwigs_setup}).
    \item[3.] The next step is a linear fit between the unsmoothed magnetic moments $\mathrm{m}_i^{\mathrm{WS}}$ and the smoothed magnetic moments $\mathrm{m}_{i}^{\mathrm{sWS}}$, the result of which is used to adapt the initial target magnetic moments $\mathrm{m}_{i}^{\mathrm{target}}$. Due to the smoothing function $F_i(|\mathbf{r}|)$, the constrained magnetic moments $\mathrm{m}_{i}^{\mathrm{sWS}}$ always differ from $\mathrm{m}_{i}^{\mathrm{target}}$, which are required to retrain mMTP. Therefore, a scaling of the constrained magnetic moments is necessary. In this work, we use a linear regression between $\mathrm{m}_i^{\mathrm{sWS}}$ and $\mathrm{m}_i^{\mathrm{WS}}$ to determine the initial constraining smoothed magnetic moments $\mathrm{m}_i^{\mathrm{constr}}$. We do not fit $\mathrm{m}_i^{\mathrm{sWS}}$ against $\mathrm{m}_i$ because, in general, the corresponding $R^2$ value is lower and the linear relation still depends weakly on $\lambda$. For more details on the linear fitting stage, see subsection~\ref{sec:fitting_stage_method}.
    \item[4.] Once we found $\mathrm{m}_i^{\mathrm{constr}}$, non-collinear soft-constrained calculations are performed during which the functional is minimized:
    \begin{equation} \label{eq:dft_en}
        E^{\text{DFT}} = \min_{\rho} \left(E[\rho; \mathbf{R}, Z, L]+\lambda \sum_{i=1}^{N_{\rm atoms}}\left({\rm m}_{i}^{\mathrm{sWS}} - {\rm m}_i^{\rm constr}\right)^2 \right).
    \end{equation}

    After minimization, the convergence criteria for magnetic moments are checked (see subsection~\ref{sec:magmom_convergence}). If at least one of the criteria is violated, the $\lambda$ value increases and the functional in Eq.~\eqref{eq:dft_en} is minimized again, starting from the density and, therefore, the magnetic moments obtained in the previous step. If all the convergence criteria are satisfied at some $\lambda$ then this configuration is considered as converged and its magnetic moments can be refined. If, after iterating over all available $\lambda$, not all criteria are met, the calculation using the lambda with the smallest mean absolute error (MAE) between ${\rm m}_{i}^{\mathrm{sWS}}$ and ${\rm m}_{i}^{\rm constr}$ is selected. A refinement is then always performed, after which the error may become acceptable even without convergence at the $\lambda$ iteration stage.

    \item[5.] Finally, magnetic moments are refined. After the soft‑constrained calculations, we obtain $\mathrm{m}_i^{\mathrm{sWS}} \simeq \mathrm{m}_{i}^{\mathrm{constr}}$. However, the magnetic moments $\mathrm{m}_{i}$ reported in the OUTCAR file slightly differ from the integrated magnetic moment $\mathrm{m}_{i}^{\mathrm{WS}}$ in the OSZICAR file due to different integration schemes. The values $\mathrm{m}_i$ are obtained by integration over PAW regions, while the $\mathrm{m}_i^{\mathrm{WS}}$ magnetic moments are computed by radial integration over atom-centered spheres with Wigner-Seitz radii. To eliminate this discrepancy, we adjust the refined constraint moments to $\widetilde{\mathrm{m}}_i^{\mathrm{refined}}$ so that, after the final calculation, $\mathrm{m}_i \simeq \mathrm{m}_i^{\mathrm{target}}$. For more details, see subsection~\ref{sec:refinement}.

\end{itemize}

\subsection{RWIGS setup}\label{sec:rwigs_setup}

Since the RWIGS tag is essential for computing $\mathrm{m}_{i}^{\mathrm{WS}}$ ($\mathrm{M}_{\mathrm{int}}$) and $\mathrm{m}_{i}^{\mathrm{sWS}}$ ($\mathrm{MW}_{\mathrm{int}}$), we cannot avoid using it, even though magnetic moments $\mathrm{m}_{i}$ in OUTCAR can be calculated without this parameter (setting the tag $\mathrm{LORBIT>11}$). In this work, we take the initial RWIGS values from the POTCAR files and perform non-collinear calculations to determine the percentage of the simulation cell covered by the atom‑centered spheres. Subsequently, we proportionally rescale the Wigner-Seitz radii so that the spheres cover almost 100~\% of the cell volume:

\begin{equation}
    \label{eq:ws_scaling}
    R_{z_i}^{\mathrm{WS}}(\mathrm{new}) = R_{z_i}^{\mathrm{WS}}(\mathrm{old}) \, \frac{1}{C^{1/3}},
\end{equation}
where $R_{z_i}^{\mathrm{WS}}(\mathrm{new})$ is the adjusted Wigner-Seitz radius for species $z_i$, $R_{z_i}^{\mathrm{WS}}(\mathrm{old})$ is the initial Wigner-Seitz radius taken from the POTCAR file, and $C$ is the fraction of the simulation cell volume originally covered by the atom-centered spheres.

If we choose RWIGS radii for all species such that almost 100~\% of the simulated cell is covered by atom‑centered spheres, the magnetic moments reported in OUTCAR ($\mathrm{m}_{i}$) and the unsmoothed moments $\mathrm{M}_{\mathrm{int}}$ become nearly identical. We carry out further calculations with these RWIGS.

\subsection{Linear fitting}\label{sec:fitting_stage_method}

In VASP, it is possible to constrain only the smoothed magnetic moments $\mathrm{m}_{i}^{\mathrm{sWS}}$, but $\mathrm{m}_{i}^{\mathrm{sWS}} \ne \mathrm{m}_{i}^{\mathrm{target}}$ because $\mathrm{m}_{i}^{\mathrm{sWS}}$ are computed with the smoothing function $F_i(|\mathbf{r}|)$ during integration. We therefore aim to obtain reasonable constraining magnetic moments $\mathrm{m}_{i}^{\mathrm{constr}}$.

To address the problem, we perform a linear fitting between $\mathrm{m}_{i}^{\mathrm{WS}} $ and $\mathrm{m}_{i}^{\mathrm{sWS}} $ as follows:

\begin{equation} \label{eq:linear_interpolation}
    \mathrm{m}_{i}^{\mathrm{WS}} = k_{\mathrm{fit}}(\lambda)\mathrm{m}_{i}^{\mathrm{sWS}}  + b_{\mathrm{fit}}(\lambda),
\end{equation}
where $k_{\mathrm{fit}}$ and $b_{\mathrm{fit}}$ are coefficients of linear fitting.

\begin{figure}[H]
\begin{center}
\includegraphics[width=0.45\columnwidth]{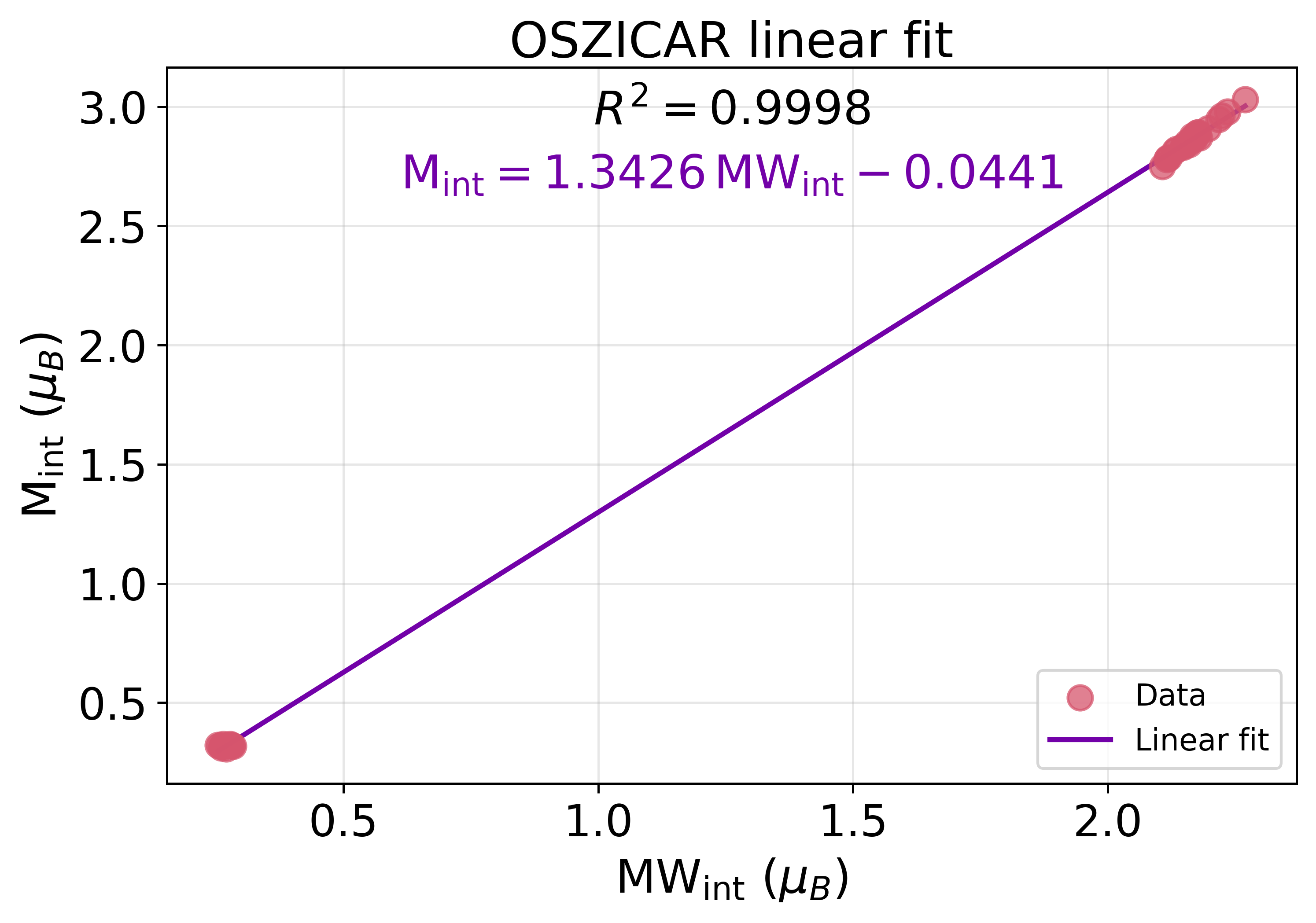}
\caption{Atom-centered integrated magnetic density at each atomic site $\mathrm{m}_{i}^{\mathrm{WS}}$ ($\mathrm{M_{int}}$) as a function of smoothed integrated magnetic density $\mathrm{m}_{i}^{\mathrm{sWS}}$ ($\mathrm{MW_{int}}$) which has been smoothed towards the boundary of the sphere. The data were linearly fitted to find coefficients for further interpolation.}\label{fig:moments_fitted_raw}
\end{center}
\end{figure}

The results of the linear fitting are shown in Fig.~\ref{fig:moments_fitted_raw}.

\begin{figure}[H]
\begin{center}
\includegraphics[width=0.5\columnwidth]{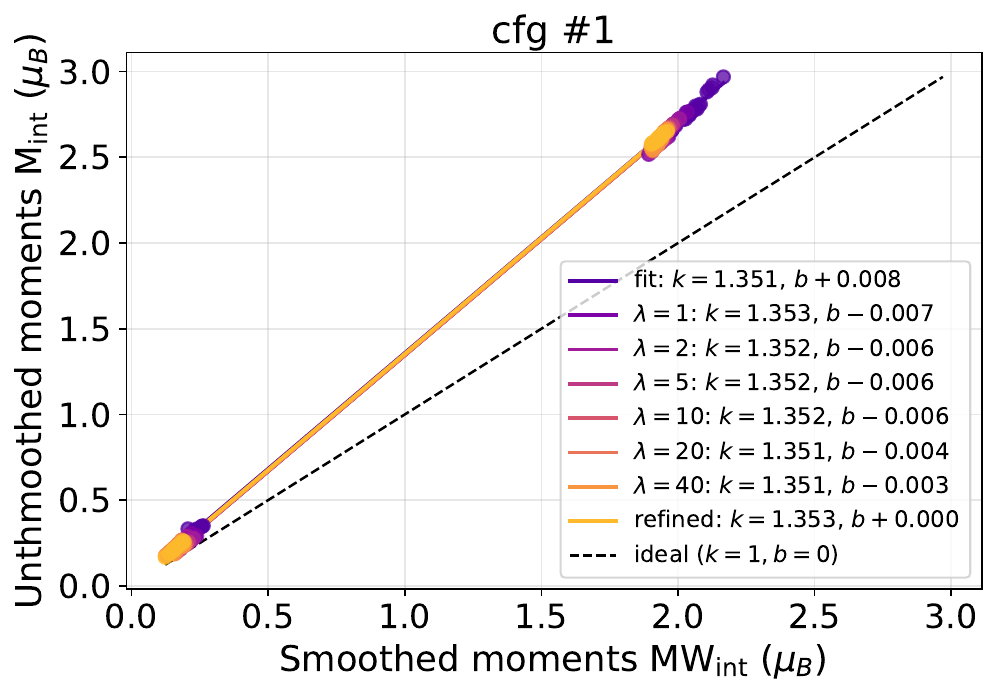}
\caption{Unsmoothed magnetic moments integrated over atom-centered spheres with Wigner-Seitz radii, $\mathrm{m}_i^{\mathrm{WS}}$ ($\mathrm{M_{int}}$), as a function of the smoothed magnetic moments $\mathrm{m}_i^{\mathrm{sWS}}$ ($\mathrm{MW_{int}}$). The interpolation coefficients remain nearly constant as the penalty weight $\lambda$ is increased.}
\label{fig:moments_fitted}
\end{center}
\end{figure}

In general, the fit parameters $k_{\mathrm{fit}}$ and $b_{\mathrm{fit}}$ depend on the penalty weight $\lambda$. However, as shown in Fig.~\ref{fig:moments_fitted}, this dependence is weak. After iterating over a sequence of penalty weights $\boldsymbol{\lambda}$, we obtain $\mathrm{m}_{i}^{\mathrm{sWS}} \simeq \mathrm{m}_{i}^{\mathrm{constr}}$, and the final refinement step removes the remaining small discrepancy. Henceforth, we use only the values $k_{\mathrm{fit}} = k_{\mathrm{fit}}(\lambda_{0})$ and $b_{\mathrm{fit}} = b_{\mathrm{fit}}(\lambda_{0})$ obtained at the first lambda weight $\lambda_{0}$.

Finally, we can get an initial approximation of constraining moments $\mathrm{m}_{i}^{\mathrm{constr}}$:

\begin{equation} \label{eq:magmom_constr}
    \mathrm{m}_{i}^{\mathrm{constr}} =  \bigl( \mathrm{m}_{i}^{\mathrm{target}} - b_{\mathrm{fit}} \bigr)/k_{\mathrm{fit}}.
\end{equation}

\subsection{Convergence criteria}\label{sec:magmom_convergence}

During iterative soft-constrained self-consistent field calculations, the gradual increase of the $\lambda$ value (see Eq.~\eqref{eq:dft_en}) is stopped and configurations are considered as converged when all the following criteria are met:

\begin{equation} 
\label{eq:max_diff_magmom}
    \max_{i} \left|{\rm m}_{i}^{\rm sWS} - {\rm m}_i^{\rm constr}\right| < \Delta M_{\rm max},
\end{equation}
where $\Delta M_{\rm max}$ is the threshold for the maximum absolute error (MaxAE),

\begin{equation} \label{eq:av_diff_magmom}
  \frac{1}{N_\mathrm{atoms}} \sum_{i=1}^{N_\mathrm{atoms}} \left| \mathrm{m}_{i}^{\mathrm{sWS}} - \mathrm{m}_i^{\mathrm{constr}} \right| < \Delta M_\mathrm{av},
\end{equation}
where $\Delta M_{\rm av}$ is the threshold for MAE,

\begin{equation} \label{eq:penalty_en}
    \lambda \sum_{i=1}^{N_{\rm atoms}}\left(\mathrm{m}_{i}^{\mathrm{sWS}} - \mathrm{m}_{i}^{\mathrm{constr}} \right)^2 /N_{\mathrm{atoms}} < E_{p}^{\mathrm{tol}}, 
\end{equation}
where $E_{p}^{\mathrm{tol}}$ is the maximum permissible contribution to the total energy of the penalty functional. The values of $\Delta M_{\rm max}$, $\Delta M_{\rm av}$, and $E_{p}^{\mathrm{tol}}$ for Fe--Pd investigated here are given in subsection~\ref{sec:soft_constr_parameters}. Details of the iterative soft-constrained self-consistent field calculations for several configurations of Fe--Pd are shown in subsection~\ref{sec:iterative_scf}. Configurations satisfying all criteria proceed to the refinement stage. Otherwise, we check smoothed magnetic moments $\mathrm{m}_i^{\rm{sWS}}$ obtained for different $\lambda$, choose the ones yielding the smallest MAE in Eq.~\eqref{eq:av_diff_magmom}, and send them to the refinement stage.

\subsection{Final refinement of magnetic moments}\label{sec:refinement}

To resolve the constant bias between the magnetic moments in OSZICAR ($\mathrm{m}_i^{\mathrm{WS}}$) and OUTCAR ($\mathrm{m}_i$) arising from different integration schemes and integration volumes, we refine the M\_CONSTR ($\mathrm{m}_i^{\mathrm{constr}}$) values in INCAR file. This consists of adjusting the constrained moments M\_CONSTR ($\mathrm{m}_i^{\mathrm{constr}}$) and rerunning the final calculation using the penalty weight $\lambda$ obtained at the end of the iterative soft-constrained stage with the refined magnetic moments $\mathrm{\widetilde{m}}^{\mathrm{refined}}_{i}$ (see Fig.~\ref{fig:scheme_soft}).

The correction proceeds as follows:

\begin{equation} \label{eq:magmom_refined}
    \mathrm{\widetilde{m}}^{\mathrm{refined}}_{i} =  \mathrm{m}^{\mathrm{constr}}_{i} - \left[ \mathrm{m}_{i}-\mathrm{m}_{i}^{\mathrm{WS}} \right] / k_{\mathrm{fit}},
\end{equation}
where $k_{\mathrm{fit}}$ is the linear slope from the fitting of $\mathrm{m}_{i}^{\mathrm{WS}}$ and $\mathrm{m}_{i}^{\mathrm{sWS}}$.

In VASP notations the correction is:

\begin{equation}
    \mathrm{M\_CONSTR(new)} = \mathrm{M\_CONSTR(old)} - \left[\mathrm{M(OUTCAR)}-\mathrm{M_{int}(OSZICAR)}\right] / k_{\mathrm{fit}},
\end{equation}
where $\mathrm{M\_CONSTR(old)}$ and $\mathrm{M\_CONSTR(new)}$ are the constrained magnetic moments in INCAR file before and after the refinement procedure, $\mathrm{M(OUTCAR)}$ are magnetic moments in OUTCAR file, $\mathrm{M(OSZICAR)}$ are magnetic moments in the OSZICAR file ($\mathrm{M}_{\mathrm{int}}$), and $k_{\mathrm{fit}}$ is the coefficient of linear interpolation between $\mathrm{M}_{\mathrm{int}}$ and $\mathrm{MW}_{\mathrm{int}}$ (see subsection~\ref{sec:fitting_stage_method}). After substituting Eqs.~\eqref{eq:linear_interpolation} and~\eqref{eq:magmom_constr} into~Eq.~\eqref{eq:magmom_refined} it can be seen that after soft-constrained DFT calculation with $\mathrm{\widetilde{m}}^{\mathrm{refined}}_{i}$ we obtain $\mathrm{m}_i \simeq \mathrm{m}_i^{\rm target}$ which was our aim.  The results of the refinement for several Fe--Pd configurations are demonstrated in subsection~\ref{sec:refinement_example}. If, after refinement, the same MaxAE, MAE, and penalty-energy thresholds as in Eqs.~\eqref{eq:max_diff_magmom},\,\eqref{eq:av_diff_magmom}, and~\eqref{eq:penalty_en} are satisfied for the differences between $\mathrm{m}_i$ and $\mathrm{m}_i^{\rm target}$ (rather than between $\mathrm{m}_i^{\mathrm{sWS}}$ and $\mathrm{m}_i^{\mathrm{constr}}$), then the configuration is added to the training set; otherwise, it is discarded.

\subsection{Parameters of soft-constrained DFT calculations for Fe--Pd}\label{sec:soft_constr_parameters}

The typical penalty weights and criteria thresholds, which we used for Fe--Pd including 32~atoms, are provided in Table~\ref{tab:soft_const_values}. The threshold for the penalty functional contribution to the total energy ($E_p^{\mathrm{tol}}$), introduced in Eq.~\eqref{eq:penalty_en}, was set below the typical ML potential training error of 1~meV/atom.

\begin{table}[H]
\centering
\begingroup
\renewcommand{\arraystretch}{1.2}
\caption{Parameters used in the Fe--Pd soft-constrained calculations and their values: maximum difference between constrained and computed magnetic moments ($\Delta M_{\mathrm{max}}$), mean absolute difference between computed and constraining magnetic moments ($\Delta M_{\mathrm{av}}$), maximum allowed penalty energy contribution to the total energy ($E_{\mathrm{p}}^{\mathrm{tol}}$), value of the penalty weight at the zero-shot iteration ($\lambda_{0}$), which is also used for Wigner-Seitz radii adjustment (see subsection~\ref{sec:rwigs_setup}) and for the fitting stage of $\mathrm{m}_i^{\mathrm{WS}}$ and $\mathrm{m}_i^{\mathrm{sWS}}$ (see subsection~\ref{sec:fitting_stage_method}), and the sequence of penalty weights employed to converge toward the desired magnetic moments ($\boldsymbol{\lambda}$). }
\label{tab:soft_const_values}
\setlength{\tabcolsep}{12pt}
\begin{tabular*}{\columnwidth}{lll}
\toprule
Parameter & Description & Value \\ \midrule
$\Delta M_{\mathrm{max}}$     & Maximum magnetic moment tolerance (MaxAE)    & $3\times10^{-2}~\mu_{\mathrm{B}}$   \\
$\Delta M_{\mathrm{av}}$     & Average magnetic moment tolerance (MAE) &  $2\times10^{-2}~\mu_{\mathrm{B}}$  \\
$E_{\mathrm{p}}^{\mathrm{tol}}$     &  Contribution to the total energy arising from the penalty functional    & $0.3$~meV/atom   \\
 $\lambda_{0}$     &  Penalty weight at the zero-shot non-collinear iterations   & 0.01    \\
 $\boldsymbol{\lambda}$    &  Penalty weights used to iteratively converge to the desired moments    & $[1,2,5,10,20,40]$   \\ \bottomrule
\end{tabular*}
\endgroup
\end{table}

\subsection{Results of soft-constrained DFT calculations}

As we have already used the maximum absolute error (MaxAE) and the mean absolute error (MAE) as convergence criteria, we decided not to employ the root mean square error (RMSE) as a stopping criterion in our workflow, because it can be excessive for practical use. However, RMSE remains useful during post‑processing analysis of the results, and therefore we introduce it as follows:

\begin{equation*}
     \Delta M_{\mathrm{RMSE}}  = \sqrt{\sum_{i=1}^{N_{\rm atoms}} ({\rm m}_{i}^{\mathrm{sWS}}-{\rm m}_i^{\rm constr})^2/N_{\rm atoms}}.
\end{equation*}

\subsubsection{Iterative soft-constrained self-consistent field calculations stage} \label{sec:iterative_scf}

\begin{figure}[H]
\begin{center}
\includegraphics[width=0.89\columnwidth]{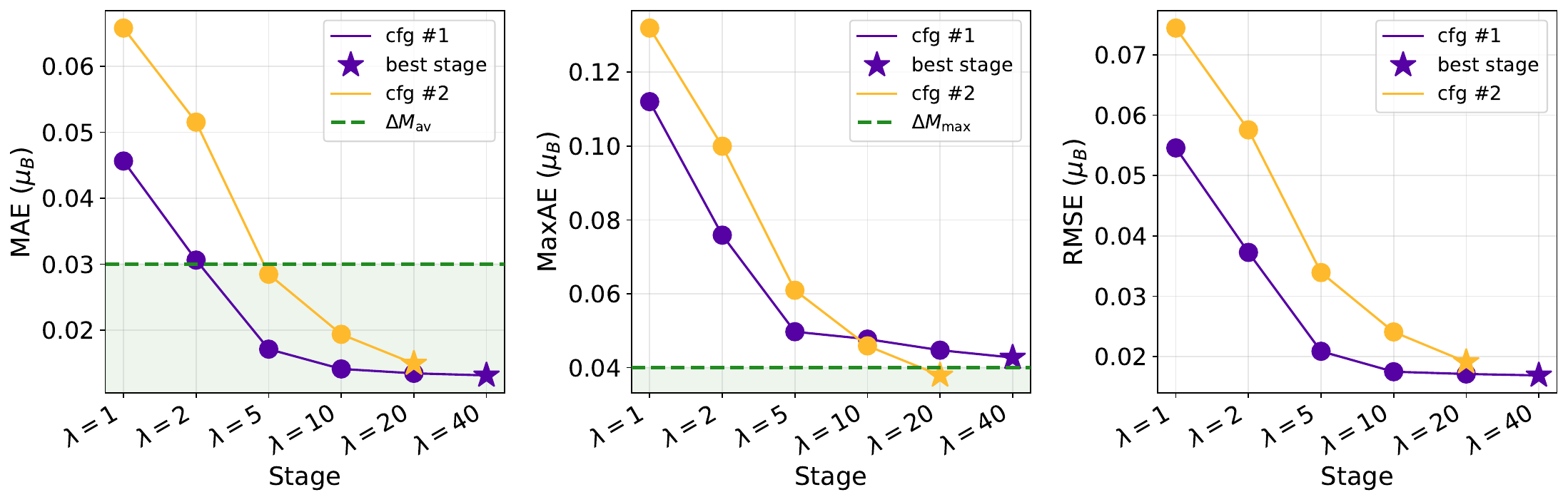}
\caption{Evolution of the error metrics (MAE, MaxAE, and RMSE) at stages with different penalty weights $\lambda$ (subsection~\ref{sec:magmom_convergence}). The errors are defined as the difference between the constraining magnetic moments $\mathrm{m}_i^{\mathrm{constr}}$ (M\_CONSTR in INCAR) and the calculated smoothed magnetic moments $\mathrm{m}_i^{\mathrm{sWS}}$ ($\mathrm{MW}_{\mathrm{int}}$ in OSZICAR). The star symbol indicates the best MAE across all calculations. The green shaded area satisfies the prescribed threshold criteria for MAE ($\Delta M_{\mathrm{av}}$) and MaxAE ($\Delta M_{\mathrm{max}}$). For the first configuration, no calculation with $\lambda = 40$ is shown, because all convergence criteria for MAE and MaxAE were already met at the previous stage with $\lambda = 20$.}
\label{fig:moments_error_evol}
\end{center}
\end{figure}

\begin{figure}[H]
\begin{center}
\includegraphics[width=0.55\columnwidth]{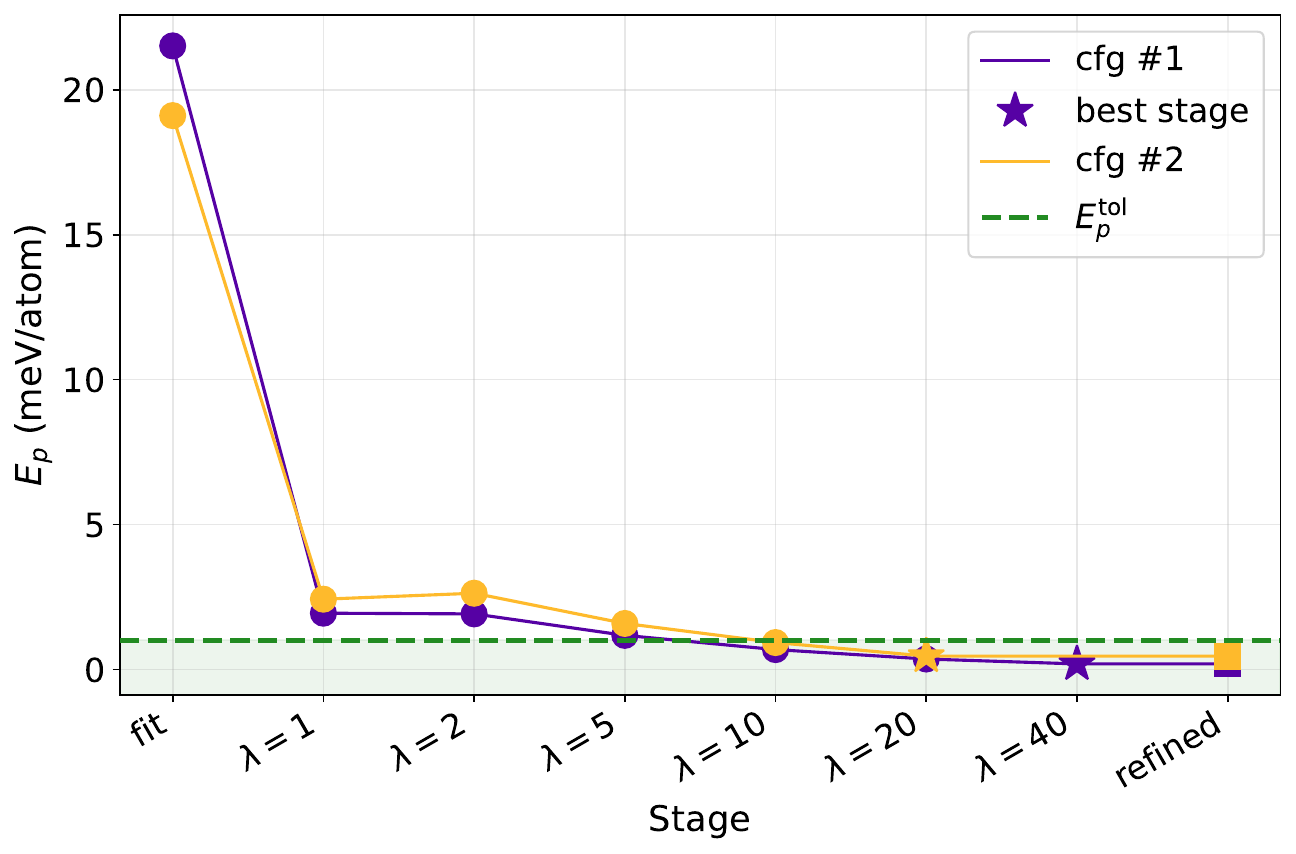}
\caption{Evolution of the penalty energy $E_{\mathrm{p}}$ across various stages for two randomly chosen configurations: fitting of magnetic moments (subsection~\ref{sec:fitting_stage_method}), convergence with different penalty weights $\lambda$ (subsection~\ref{sec:magmom_convergence}), and the refinement stage (subsection~\ref{sec:refinement}). The star symbol marks the stage with the best MAE across all considered configurations. The green shaded area indicates where the penalty energy satisfies the threshold criterion $E_{\mathrm{p}}^{\mathrm{tol}}$.}
\label{fig:moments_penalty_evol}
\end{center}
\end{figure}

Here we provide the results of the iterative soft-constrained self-consistent field calculations stage. From Figs.~\ref{fig:moments_error_evol} and~\ref{fig:moments_penalty_evol} we conclude that the convergence criteria in Eqs.~\eqref{eq:max_diff_magmom},\,\eqref{eq:av_diff_magmom}, and~\eqref{eq:penalty_en} are valid for two given configurations.

\subsubsection{Refinement stage} \label{sec:refinement_example}

Here, we replace the soft‑constraint criteria for magnetic moments ($M_{\mathrm{av}}$ and $M_{\mathrm{max}}$), which were originally defined between the constraining moments $\mathrm{m}_{i}^{\mathrm{constr}}$ and the smoothed moments $\mathrm{m}_{i}^{\mathrm{sWS}}$, with new criteria based on the difference between the calculated magnetic moments $\mathrm{m}_{i}$ and the target magnetic moments $\mathrm{m}_{i}^{\mathrm{target}}$. These new thresholds are denoted as $\delta m_{\mathrm{tol}}^{\mathrm{av}}$ for MAE and $\delta m_{\mathrm{tol}}^{\mathrm{max}}$ for MaxAE.

\begin{figure}[H]
\begin{center}
\includegraphics[width=0.85\columnwidth]{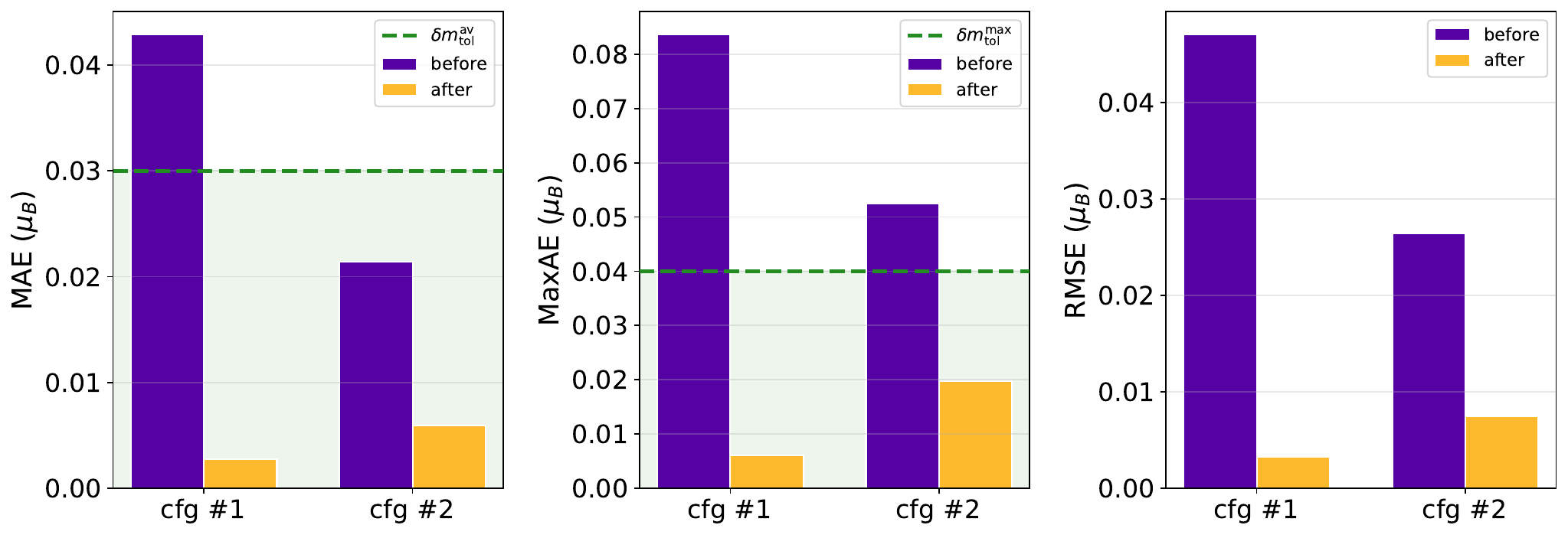}\caption{
Error metrics (MAE, MaxAE, and RMSE) before and after refinement. Here, the errors are defined as the difference between the target magnetic moments $\mathrm{m}_{i}^{\mathrm{target}}$ and the calculated magnetic moments $\mathrm{m}_{i}$ (from OUTCAR). The green shaded area indicates where both MaxAE and MAE satisfy the respective threshold criteria $\delta m_{\mathrm{tol}}^{\mathrm{max}}$ and $\delta m_{\mathrm{tol}}^{\mathrm{av}}$.
}
\label{fig:refinement_metrics}
\end{center}
\end{figure}

\begin{figure}[H]
\begin{center}
\includegraphics[width=0.90\columnwidth]{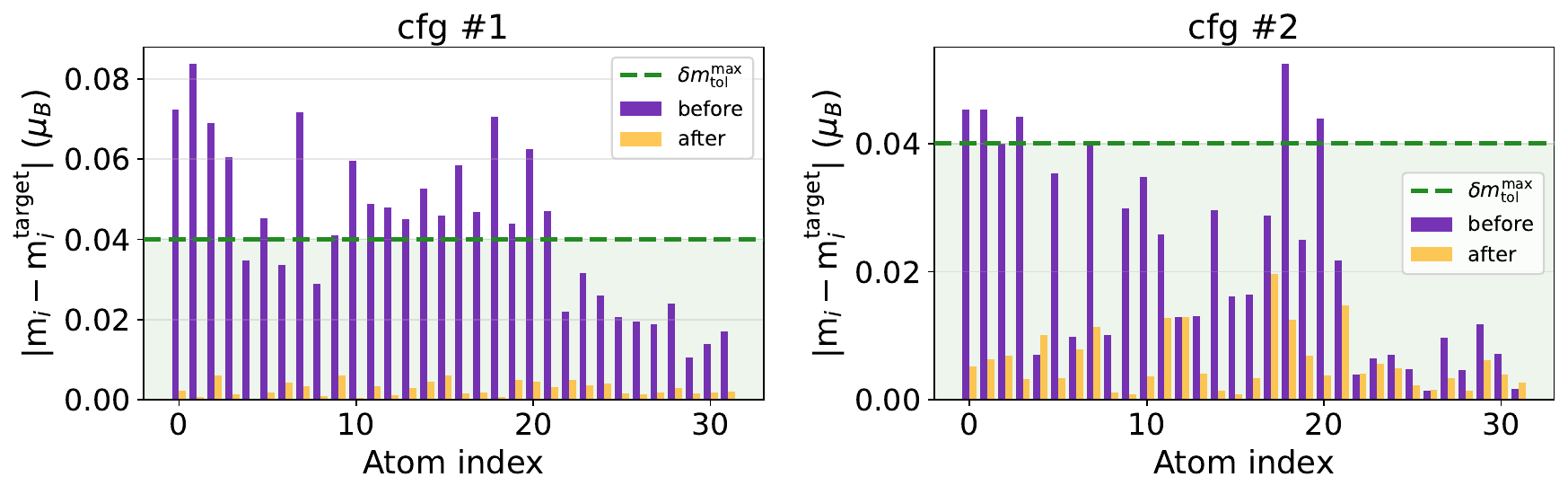}\caption{
Per‑atom results of the magnetic‑moment refinement for two configurations. Here, we compare the target magnetic moments $\mathrm{m}_{i}^{\mathrm{target}}$ with the calculated magnetic moments $\mathrm{m}_{i}$ (from OUTCAR), which are used in the mMTP training. (left) first configuration; (right) second configuration. The green shaded area indicates where MaxAE satisfies the respective threshold $\delta m_{\mathrm{tol}}^{\mathrm{max}}$.
}
\label{fig:refinement_per_atom}
\end{center}
\end{figure}

After refinement, the difference between the calculated magnetic moments $\mathrm{m}_{i}$ and the target moments $\mathrm{m}_{i}^{\mathrm{target}}$ decreases both on average (Fig.~\ref{fig:refinement_metrics}) and for each individual atom (Fig.~\ref{fig:refinement_per_atom}).

\subsection{Convergence of magnetic moments}

\begin{figure}[H]
\begin{center}
\includegraphics[width=0.45\columnwidth]{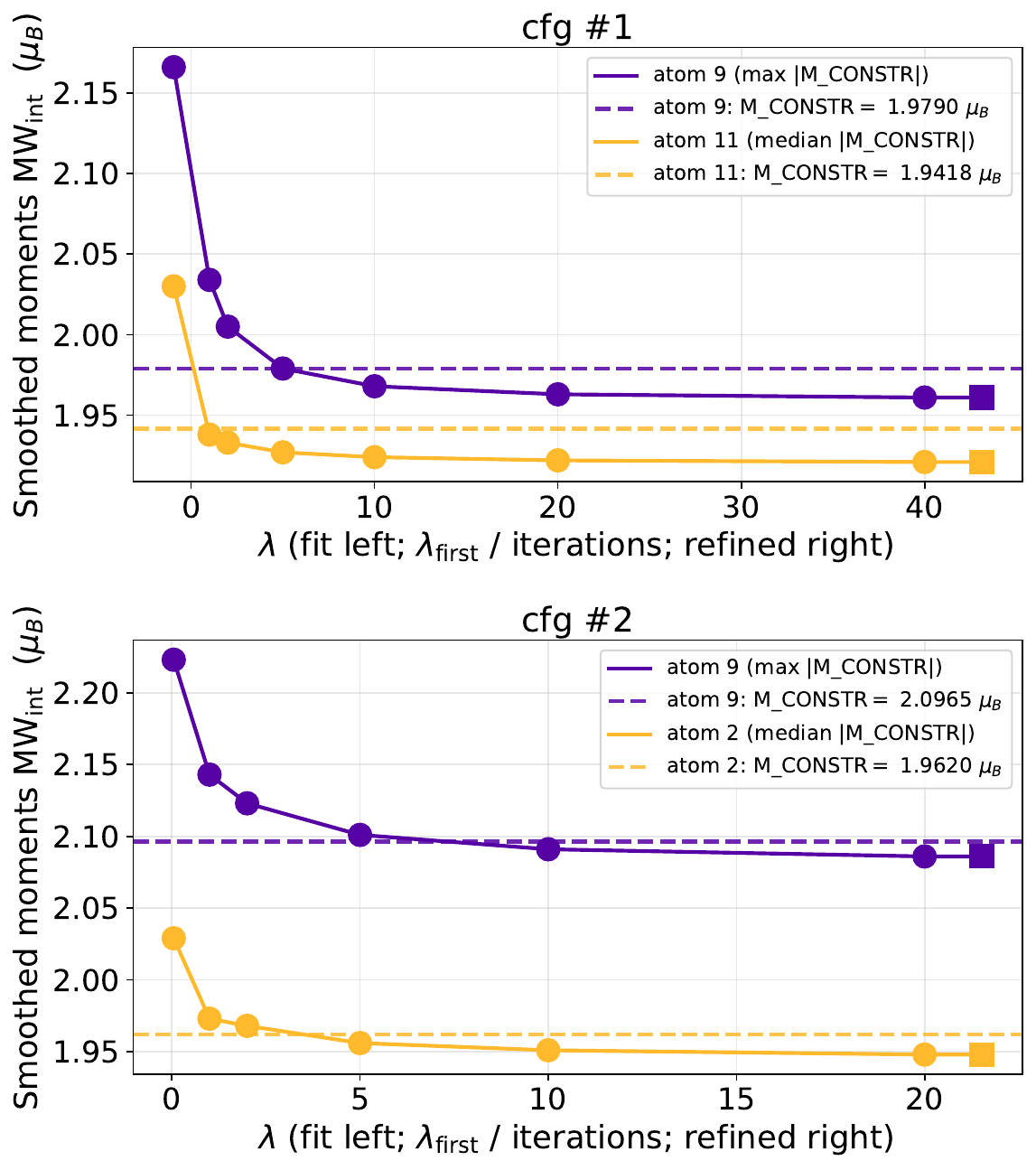}
\caption{
Smoothed magnetic moments $\mathrm{m}_i^{\mathrm{sWS}}$ (reported in OSZICAR) at various stages of the workflow, illustrated with solid lines: the fitting stage, the series of soft-constrained calculations with increasing penalty weight $\lambda$, and the refinement stage. Dashed lines show the constraining magnetic moments $\mathrm{m}_i^{\mathrm{constr}}$ (M\_CONSTR). The purple solid line corresponds to the maximum absolute target moment, and the yellow solid line to the median of the absolute values over all target moments. Squares denote the results from the refinement stage, while circles correspond to the other stages. We show results for two randomly selected configurations: (top) the first configuration and (bottom) the second configuration. 
}
\label{fig:conv_mw_int_to_m_constr}
\end{center}
\end{figure}

\begin{figure}[H]
\begin{center}
\includegraphics[width=0.45\columnwidth]{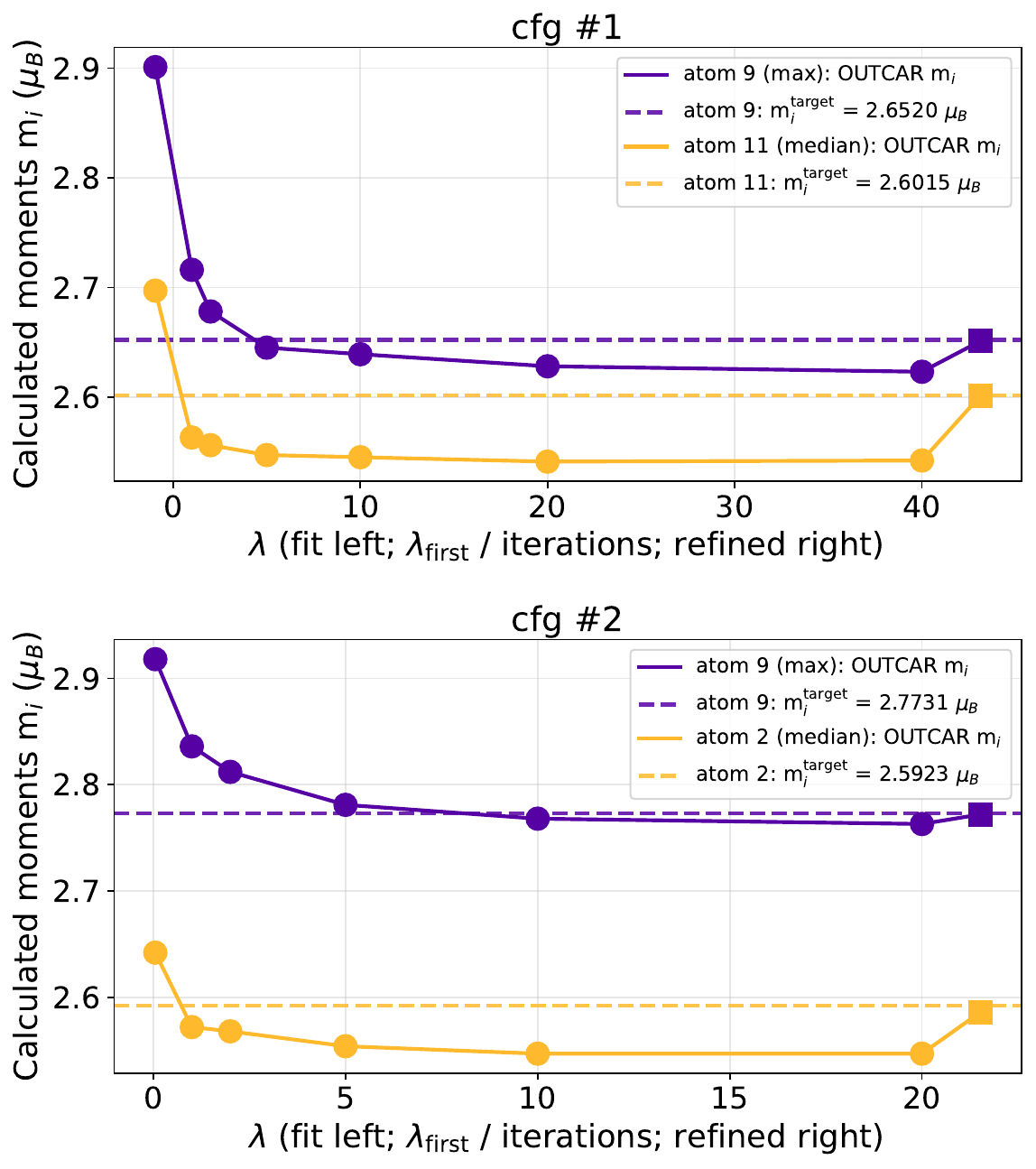}
\caption{
Calculated magnetic moments $\mathrm{m}_i$ (reported in OUTCAR) at various stages of the workflow, illustrated with solid lines: the fitting stage, the series of soft-constrained calculations with increasing penalty weight $\lambda$, and the refinement stage. Dashed lines show the target magnetic moments $\mathrm{m}_i^{\mathrm{target}}$. The purple solid line corresponds to the maximum absolute target moment, and the yellow solid line to the median of the absolute values over all target moments. Squares denote the results from the refinement stage, while circles correspond to the other stages. We show results for two randomly selected configurations: (top) the first configuration and (bottom) the second configuration. 
}
\label{fig:conv_outcar_to_m_constr}
\end{center}
\end{figure}

Summarizing our workflow, the soft‑constrained calculations allow us to obtain the smoothed magnetic moments $\mathrm{m}_i^{\mathrm{sWS}}$, as illustrated in Fig.~\ref{fig:conv_mw_int_to_m_constr}. However, due to the different integration schemes used for $\mathrm{m}_i$ and $\mathrm{m}_i^{\mathrm{WS}}$, where the latter is linearly fitted to $\mathrm{m}_i^{\mathrm{sWS}}$, a small discrepancy remains between the target magnetic moments $\mathrm{m}_i^{\mathrm{target}}$ and the final moments $\mathrm{m}_i$ (reported in OUTCAR), as shown in Fig.~\ref{fig:conv_outcar_to_m_constr}. This discrepancy is successfully eliminated after the refinement stage. We obtain $\mathrm{m}_i \simeq \mathrm{m}_{i}^{\mathrm{target}}$, and all magnetic moments satisfy the required criteria for MaxAE and MAE.

\section{Timing benchmarks of soft-constrained calculations}\label{sec:timings_si_soft}

To validate the efficiency of our soft-constrained pipeline, we randomly selected five configurations from the pool of selected configurations generated during active learning and benchmarked the computational cost of each stage. The elapsed time of VASP calculations was used as the performance metric. If the convergence criteria were met with $\lambda_i < \lambda_{\mathrm{max}}$, the refinement stage was subsequently performed. All timings were normalized by the corresponding collinear SCF calculation performed with equilibrium moments for each configuration. To minimize performance variability, all test calculations were executed on 16 cores of an AMD EPYC 9634 central processing unit (CPU).

\begin{figure}[H]
\begin{center}
\includegraphics[width=0.52\columnwidth]{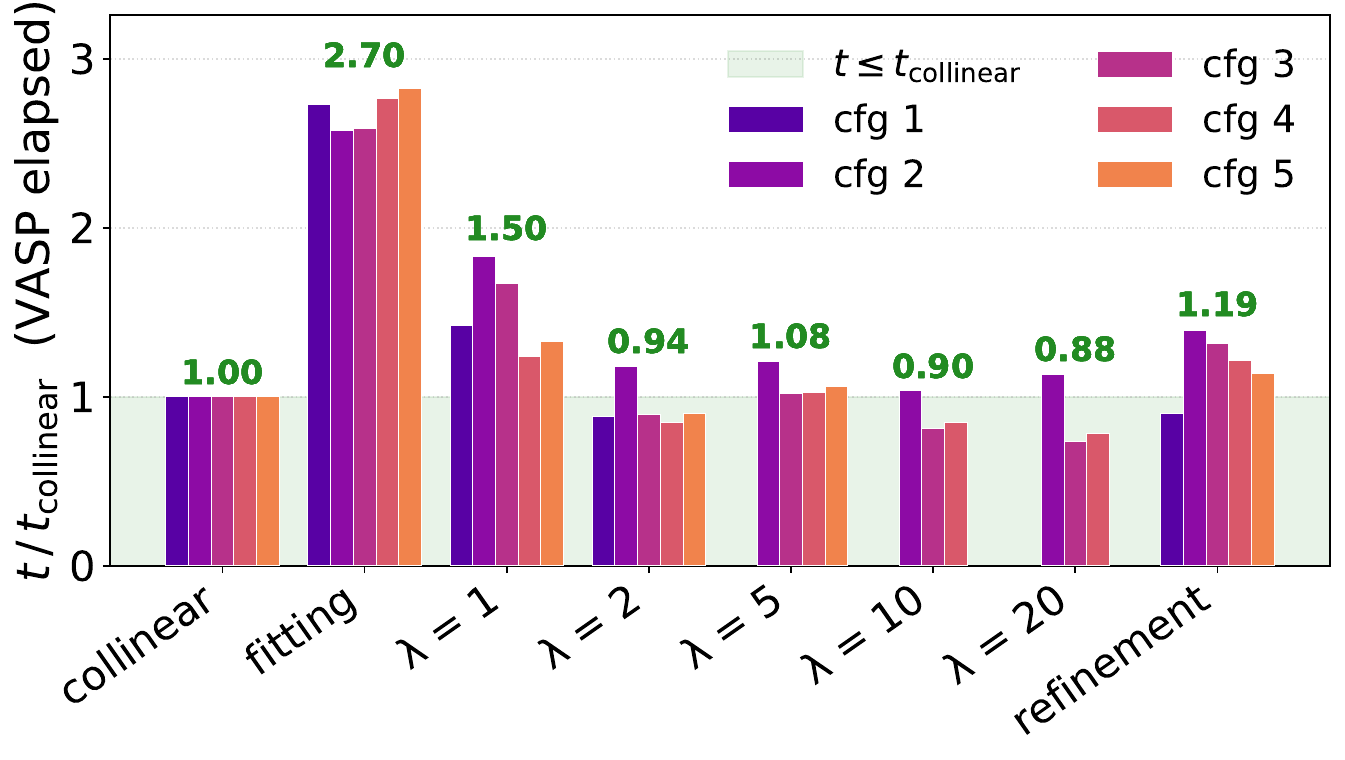}
\caption{Bar chart for the normalized timing breakdown of our soft-constrained pipeline. Five configurations were randomly sampled from the pool of selected configurations, and for each, we report the computational cost of the fitting stage, the soft-constraint iterations (with $\lambda_i$), and the refinement stage. All timings are normalized by the cost of a single collinear SCF calculation on the same configuration. The green text above each group of bars indicates the average time for that stage. The semi-transparent green region marks calculations that were faster than the corresponding collinear calculation for each configuration.
}
\label{fig:timings_norm_bars}
\end{center}
\end{figure}

The timing breakdown is presented as a bar graph in Fig.~\ref{fig:timings_norm_bars}.
The results indicate that the fitting stage is the most expensive step during non-collinear calculations, as no initial charge density (CHGCAR) or wave function (WAVECAR) is available for the configuration at this point. Consequently, the cost of subsequent non-collinear calculations decreases substantially, owing to the improved starting guess provided by the previously converged CHGCAR and WAVECAR files.

\begin{figure}[H]
\begin{center}
\includegraphics[width=0.55\columnwidth]{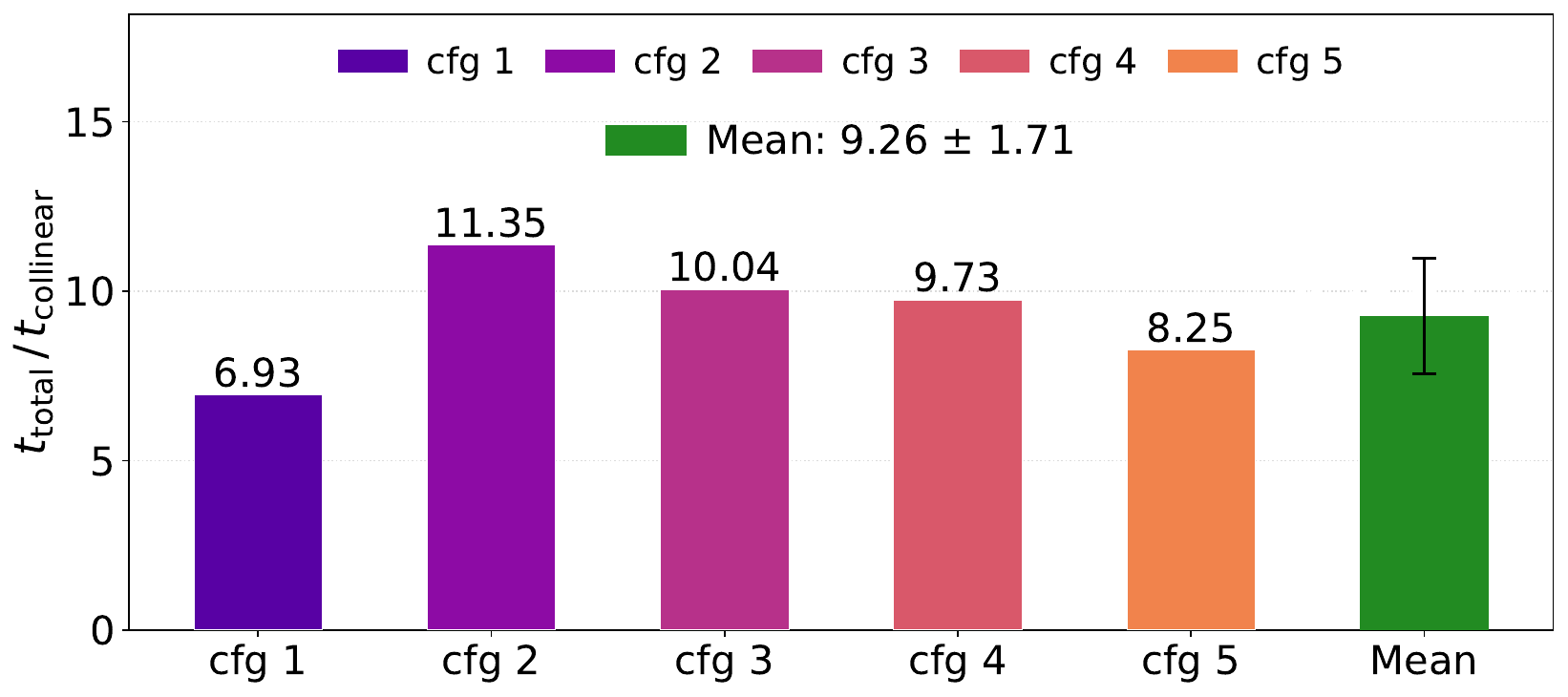}
\caption{Total computational cost of the soft-constrained pipeline for five randomly sampled configurations from the pool of selected configurations. For each configuration, the total time of the complete soft-constrained calculation is shown, normalized by the cost of a single collinear SCF calculation on the same configuration. The green bar indicates the mean value with its standard deviation.}

\label{fig:collinear_noncol_ratio_vasp}
\end{center}
\end{figure}

The ratio of total non-collinear to collinear timings is shown for each configuration in Fig.~\ref{fig:collinear_noncol_ratio_vasp} with average values provided. The corresponding real average CPU timings (in hours) across all configurations are summarized in Table~\ref{tab:dft_cpu_hours}. On average, the entire soft-constraining pipeline requires approximately $9.3 \pm 1.7$ times the computational cost of a single collinear equilibrium calculation. In this study, we selected the set of $\lambda_{i}$ to maximize the number of configurations that converge during the soft-constrained pipeline. However, a system-optimized set of penalty weights could further reduce the cost of the pipeline.

\begin{table}[H]
\centering
\caption{
CPU-hours per DFT stage, compiled over five configurations  ($N_{\mathrm{cores}}=16$; VASP elapsed time). $n$ is the number of  configurations that ran that stage. ``All $\lambda$'' and ``all non-collinear'' denote the sums of the corresponding jobs per configuration, taken before averaging. None of the configurations reached the stage with $\lambda=40$.
}
\label{tab:dft_cpu_hours}
\begin{tabular}{lcc}
\toprule
Stage & $n$ & CPU-hours \\
\midrule
Collinear & 5 & $16.69 \pm 1.02$ \\
Fitting stage & 5 & $45.08 \pm 3.87$ \\
$\lambda=1$ & 5 & $24.91 \pm 3.54$ \\
$\lambda=2$ & 5 & $15.68 \pm 2.01$ \\
$\lambda=5$ & 4 & $17.58 \pm 1.42$ \\
$\lambda=10$ & 3 & $14.63 \pm 1.98$ \\
$\lambda=20$ & 3 & $14.40 \pm 3.46$ \\
Refinement stage & 5 & $19.78 \pm 2.31$ \\
\midrule
All $\lambda$ & 5 & $72.07 \pm 24.23$ \\
All non-collinear & 5 & $136.92 \pm 23.61$ \\
Pipeline total & 5 & $153.62 \pm 23.04$ \\
\bottomrule
\end{tabular}
\end{table}

\clearpage
\section{Examples of VASP input files}

Below, we provide examples of INCAR files for both unconstrained and constrained DFT calculations in VASP, which we used for Fe--Pd.

Example 1. INCAR file for the unconstrained DFT calculations (LNONCOLLINEAR = .FALSE., LAMBDA=0).
\noindent
\begin{verbatim}
SYSTEM = equilibrium non-constrained FePd calculations
ALGO = Normal
EDIFF = 1e-06
ENCUT = 500
IBRION = 1
ICHARG = 2
ISMEAR = 1
ISPIN = 2
ISTART = 0
ISYM = 0
KGAMMA = .TRUE.
LCHARG = .TRUE.
LORBIT = 11
LSCALU = .FALSE.
LWAVE = .TRUE.
NBANDS = 300
NELM = 200
NELMIN = 4
NSW = 0
NPAR = 1
PREC = Accurate
SIGMA = 0.1
RWIGS = 1.404 1.302
\end{verbatim}

\clearpage

Example 2. INCAR file for the soft-constrained DFT calculations (LNONCOLLINEAR = .TRUE., LAMBDA$>$0).
\noindent
\begin{verbatim}
SYSTEM = non-equilibrium soft-constrained FePd calculations
ALGO = Normal
EDIFF = 1e-05
ENCUT = 500
IBRION = 1
ICHARG = 1
ISMEAR = 1
ISPIN = 2
ISTART = 1
ISYM = 0
I_CONSTRAINED_M = 2
KGAMMA = .TRUE.
LAMBDA = 0.01
LCHARG = .TRUE.
LNONCOLLINEAR = .TRUE.
LORBIT = 11
LSCALU = .FALSE.
LWAVE = .TRUE.
MAGMOM  =  2.211 0 0 2.064 0 0 2.260 0 0 2.250 0 0 2.280 0 0 2.162 0 0 2.142 0 0 2.151 0 0 2.345 0 0 
 2.216 0 0 2.286 0 0 2.261 0 0 2.087 0 0 2.294 0 0 2.220 0 0 2.233 0 0 2.220 0 0 2.195 0 0 2.296 0 0 
 2.137 0 0 2.245 0 0 2.222 0 0 0.247 0 0 0.238 0 0 0.233 0 0 0.223 0 0 0.231 0 0 0.222 0 0 0.247 0 0 
 0.225 0 0 0.245 0 0 0.240 0 0
M_CONSTR = 2.211 0 0 2.064 0 0 2.260 0 0 2.250 0 0 2.280 0 0 2.162 0 0 2.142 0 0 2.151 0 0 2.345 0 0 
 2.216 0 0 2.286 0 0 2.261 0 0 2.087 0 0 2.294 0 0 2.220 0 0 2.233 0 0 2.220 0 0 2.195 0 0 2.296 0 0 
 2.137 0 0 2.245 0 0 2.222 0 0 0.247 0 0 0.238 0 0 0.233 0 0 0.223 0 0 0.231 0 0 0.222 0 0 0.247 0 0 
 0.225 0 0 0.245 0 0 0.240 0 0
NBANDS = 850
NELM = 200
NELMIN = 4
NSW = 0
NPAR = 1
PREC = Accurate
SIGMA = 0.1
RWIGS = 1.404 1.302
\end{verbatim}

\clearpage
\section{Scaling of magnetic Moment Tensor Potential}

The motivation for applying scaling to magnetic Moment Tensor Potential during fitting is to accelerate the convergence of optimization with the BFGS algorithm. To that end, we minimize the spread ratio $\rho$ of the linear coefficients $\xi_{\alpha}$:
    \begin{equation} \label{eq:spread_ratio}
        \rho = \sqrt{\sum_{\alpha=1}^{N_{\mathrm{lin}}} \xi_{\alpha}^2}\,\big/\,\xi_{\rm med},
    \end{equation}
where $\xi_{\rm med} = \operatorname{median}\bigl(|\xi_1|, \ldots,
|\xi_{N_{\rm lin}}|\bigr)$.

The optimal scaling parameter is determined iteratively according to the following procedure.
\begin{enumerate}
    \item Set up an initial $s$ (e.g., $s=1$).
    \item Choose a set of scaling parameters $\mathbf{s} = (\tilde{s}_l)_{l=1}^{5}
          = (1.2s,\, 1.1s,\, s,\, 0.9s,\, 0.8s)$,
          where $s$ denotes the current scaling parameter in \eqref{eq:radial_part}.
    \item For each $\tilde{s}_l$, construct the system of radial vectors
          ${\bm c}_{0}, \ldots,
           {\bm c}_{\mu}, \ldots,
           {\bm c}_{\mu_{\rm max}},$
          for each radial function indexed by $\mu=0,\ldots,\mu_{\rm max}$, and orthonormalize these vectors using the Gram--Schmidt procedure. Here ${\bm c}_\mu = (c_{\mu, z_i, z_j}^{\zeta, \beta, \gamma})_{z_i, z_j, \zeta, \beta, \gamma}$.
    \item Solve the system of linear equations
    \begin{equation} \label{eq:linear_system}
        \sum_{\alpha=1}^{N_{\rm lin}} \sum_{i=1}^{N_{\rm atoms}}
        \xi_{\alpha} \, B_{\alpha}^{(k)}(\mathcal{N}_{i})
        = E_{\rm DFT}^{(k)}, \qquad k = 1, \ldots, K,
    \end{equation}
    with the fixed orthonormalized radial vectors obtained for each $s_l$.
    \item Compute the spread ratio $\rho_l$ for each $\tilde{s}_l$ and update the scaling parameter according to:
          $s = \tilde{s}_{l^*}$, where $l^* = \arg\min_l \rho_l$.
    \item Repeat steps 2--5 until $l^* = 3$ (i.e., $\tilde{s}_l=s$).
\end{enumerate}
The described procedure yields a scaling parameter $s$ that makes the linear coefficients $\xi_\alpha$ comparable in magnitude, stabilizing the linear system \eqref{eq:linear_system} and, therefore, accelerating the convergence of BFGS.

\end{document}